\documentclass[%
reprint,
superscriptaddress,
showpacs,preprintnumbers,
 amsmath,amssymb,
 aps,
prd,
]{revtex4-1}

\usepackage{float}
\usepackage{graphicx}
\usepackage{dcolumn}
\usepackage{bm}
\usepackage{bbold}
\usepackage{amssymb,amsmath}
\usepackage{hyperref}

\usepackage{color}
\usepackage{amsfonts}
\usepackage{subfigure}
\usepackage{array}

\newcommand{\Tr}{\ensuremath{\operatorname{Tr}}}

\newcolumntype{L}{>{\centering\arraybackslash}m{3cm}}

\definecolor{bjcol}{rgb}{1,.44,0.13}

\definecolor{blue}{rgb}{0,0,1}

\definecolor{green}{rgb}{0,1,0}

\definecolor{red}{rgb}{1,0,0}

\definecolor{gray}{rgb}{.5,.5,.5}

\definecolor{darkgreen}{rgb}{.0,.5,.0}

\def\Fig#1{Fig.~\ref{#1}}

\def\Eq#1{Eq.~(\ref{#1})}
\def\eq#1{(\ref{#1})}
\def\eqref#1{(\ref{#1})}

\def\sec#1{Sec.~\ref{#1}}
\def\app#1{Appendix~\ref{#1}}

\def\lA0{{\langle A_0 \rangle}}
\def\bA0{{\bar{A}_0}}

\def\0#1#2{\frac{#1}{#2}}

\graphicspath{{./figures/}{./}}

\begin{document}

\preprint{}

\title{Mesonic dynamics and QCD phase transition
}

\author{Shi Yin}
\affiliation{School of Physics, Dalian University of Technology, Dalian, 116024,
  P.R. China}

\author{Rui Wen}
\affiliation{School of Physics, Dalian University of Technology, Dalian, 116024,
  P.R. China}

\author{Wei-jie Fu}
\email{wjfu@dlut.edu.cn}
\affiliation{School of Physics, Dalian University of Technology, Dalian, 116024,
  P.R. China}


\begin{abstract}

We study the nontrivial dispersion relation of mesons resulting from the splitting of the transversal and longitudinal mesonic wave function renormalizations, and its influences on the QCD phase transition, equation of state, and the fluctuations of baryon number. The calculations are performed in a two-flavor low energy effective model within the functional renormalization group approach. We find that influences of the splitting mesonic wave function renormalizations on the equilibrium thermodynamical bulk properties are mild. Furthermore, we have compare the fixed and running expansion for the effective potential in detail, through which the role of the field-dependent meson and quark wave function renormalizations could be inferred.

\end{abstract}

\maketitle



\section{Introduction}
\label{sec:int}

Recent years have seen remarkable progresses in the studies on the QCD phase structure from both the experimental and theoretical sides. Non-monotonic behavior of non-Gaussian cumulants of the net-proton distribution have been observed in Phase I of the Beam Energy Scan (BES) Program at the Relativistic Heavy Ion Collider (RHIC) \cite{Adamczyk:2013dal,Luo:2015ewa,Luo:2017faz}. This non-monotonic behavior is potentially related to the critical end point (CEP) in the QCD phase diagram in terms of the temperature and the baryon chemical potential or the baryon density. The CEP is usually believed to be a key feature of the QCD phase structure, which separates the first order phase transition at high density from the continuous crossover at low density. 

On the theoretical side, lattice QCD simulation is the first-principle approach to explore the properties of QCD at finite temperature and densities. There have been significant progresses in the lattice studies in recent years, such as the phase boundary \cite{Bellwied:2015rza,Bazavov:2018mes}, QCD equation of state at finite baryon chemical potential and the fluctuations and correlations of the conserved charges \cite{Bazavov:2017dus,Bazavov:2017tot,Borsanyi:2018grb}, the chiral phase transition temperature \cite{Ding:2019prx}, etc. Because of the sign problem at finite chemical potential, computation of lattice QCD is still restricted to the regime of small chemical potential in the phase diagram, e.g., $\mu_B/T\le 2\sim 3$. Functional continuum field approaches, such as the Dyson-Schwinger equation \cite{Fischer:2018sdj}, the functional renormalization group (FRG) \cite{Pawlowski:2005xe}, etc., are also very suited for the studies of QCD at finite temperature and densities. Functional approaches are not hampered by the sign problem, thus can be employed to investigate the whole phase diagram \cite{Schaefer:2004en,Herbst:2013ail,Fischer:2018sdj}, in particular to predict the location of CEP, e.g., see \cite{Braun:2009gm,Fischer:2014ata,Gao:2015kea,Isserstedt:2019pgx,Fu:2019hdw}.

In this work we will employ the FRG to study QCD thermodynamics, phase transition and the baryon number fluctuations. In the FRG approach, quantum fluctuations of different scales are integrated successively, with the renormalization group (RG) scale evolving from the ultraviolet $k=\Lambda$ to the infrared $k=0$, see e.g., \cite{Berges:2000ew,Pawlowski:2005xe,Pawlowski:2010ht,Braun:2011pp} for QCD-related reviews. Remarkably, significant progresses have been made in the first-principle QCD studies within the FRG approach in the vacuum \cite{Mitter:2014wpa,Braun:2014ata,Rennecke:2015eba,Cyrol:2016tym,Cyrol:2017ewj,Cyrol:2017qkl} and at finite temperature and densities \cite{Fu:2019hdw} in recent years. Furthermore, low energy effective models within the FRG approach have provided us with a wealth of knowledges about the properties of the strongly interacting matter at finite temperature and densities, for more details, see e.g. \cite{Schaefer:2004en,Schaefer:2006sr,Skokov:2010wb,Herbst:2010rf,Schaefer:2011ex,Mitter:2013fxa,Herbst:2013ufa,Tripolt:2013jra,Fu:2015naa,Fu:2015amv,Fu:2016tey,Rennecke:2016tkm,Jung:2016yxl,Braun:2017srn,Fu:2018qsk,Fu:2018swz,Sun:2018ozp,Wen:2018nkn,Li:2019nzj,Wen:2019ruz}.

The low-energy effective models, e.g. the quark-meson (QM) model \cite{Schaefer:2004en}, Nambu--Jona-Lasinio (NJL) model \cite{Buballa:2003qv}, and their Polyakov-loop improved variants: PQM and PNJL, are suitable to be employed to study the QCD phase transitions. They have been investigated quite a lot in literatures, see, e.g., \cite{Fukushima:2003fw,Ratti:2005jh,Fu:2007xc,Schaefer:2007pw,Fu:2009wy,Fu:2010ay} for more details. In this work, we adopt the scale-dependent effective action for the two-flavor PQM model, as follows 
\begin{align}
\Gamma_k=&\int_x \bigg\{Z_{q,k}\bar{q} \Big [\gamma_\mu \partial_\mu -\gamma_0(\hat\mu+igA_0) \Big ]q \nonumber\\[2ex]
&+\frac{1}{2}\Big [Z_{\phi,k}^{\parallel}(\partial_0 \phi)^2+Z_{\phi,k}^{\perp}(\partial_i \phi)^2 \Big]\nonumber\\[2ex]
&+h_k(\rho)\bar{q}\big(T^0\sigma+i\gamma_5\vec{T}\cdot \vec{\pi}\big)q+V_k(\rho)-c\sigma \bigg\}\,,\label{eq:action}
\end{align}
with $\mu=(0, 1, ..., 3)$ and $i=(1, 2, 3)$. In \Eq{eq:action} we have used notation $\int_{x}=\int_0^{1/T}d x_0 \int d^3 x$, where the imaginary time formalism for the field theory at finite temperature is used, and the temporal length reads $\beta=1/T$. Apparently, when the temperature is nonzero, the $O(4)$-symmetry in the Euclidean space is broken into that of $\mathbb{Z}_2\otimes O(3)$, which leads to the split of the magnetic and electric components of correlations functions. They correspond to the components transversal and longitudinal to the heat bath, respectively. In this work, we take this split into account in the two-point correlation function for the mesons, as shown in the second line on the r.h.s. of \Eq{eq:action}, where $Z_{\phi,k}^{\parallel}$ and $Z_{\phi,k}^{\perp}$ indicate the longitudinal and transversal wave function renormalizations for the temporal and spatial components, respectively. 

The reason why we concentrate on the splitting of the wave function renormalization, especially for the mesons, is due to the facts as follows. Firstly, in comparison to the quark wave function renormalization $Z_{q,k}$ and the scale dependent Yukawa coupling $h_k$ in \Eq{eq:action}, it is found that the meson wave function renormalization $Z_{\phi,k}$ plays the most significant role beyond the local potential approximation (LPA) \cite{Pawlowski:2014zaa,Fu:2015naa}. In the LPA, the propagators are classical, i.e., $Z_{q,k}=Z_{\phi,k}=1$ and the Yukawa coupling $h$ is a constant and independent of the scale $k$. Secondly, In Ref. \cite{Helmboldt:2014iya} calculations based on the full momentum-dependent two-point correlation functions of mesons are compared with those from LPA and LPA$'$, where a momentum-independent $Z_{\phi,k}$ is included in LPA$'$ in comparison to LPA, and it is found that there is a good agreement between the full momentum calculation and the LPA$'$, rather not LPA, which indicates that the dispersion relation for the meson, resulting from a scale dependent $Z_{\phi,k}$, have already captured most momentum dependence of the two-point correlation function. 

Considering the importance of the wave function renormalization for the mesons and the success of LPA$'$, in this work we would like to investigate the effects of the splitting of $Z_{\phi,k}$ in the LPA$'$ as shown in \Eq{eq:action}, which is a natural choice at finite temperature as discussed above. Furthermore, we will also study the interplay between the splitting of $Z_{\phi,k}$ and other truncation approaches, e.g., the field dependent Yukawa coupling $h_k(\rho)$ which encodes higher order quark-meson scattering processes \cite{Pawlowski:2014zaa}, fixed point expansion for the effective potential $V_k(\rho)$ in \Eq{eq:action} versus the running point expansion, etc. Their influences on the QCD phase transition and observables, e.g. fluctuations of the baryon number, will be investigated in detail.

This paper is organized as follows. In \sec{sec:FRG} we give the flow equations for the effective potential, anomalous dimensions, and the Yukawa coupling. The equation of state and the baryon number fluctuations are presented in \sec{sec:EoS}. In \sec{sec:num} we show our numerical results, and finally a summary and outlook is presented in \sec{sec:sum}. Explicit expressions for the anomalous dimensions and the flow of the Yukawa are given in \app{app:anom}, and some useful threshold functions are collected in \app{app:thresholdfun}.


\section{Functional renormalization group and flow equations}
\label{sec:FRG}

To proceed, we describe other notations in the effective action in \Eq{eq:action}. $\phi=\left(\sigma,\vec{\pi}\right)$ is a meson field with four components. The effective potential $V_k(\rho)$ with $\rho=\phi^2/2$ is $O(4)$ invariant and the $c$-term $c\sigma$ breaks the chiral symmetry explicitly. The mesons interact with quarks through the scalar and pseudo-scalar channels with a mesonic field dependent Yukawa coupling $h_k(\rho)$, and $T^0$ and $T^i$ are the generators in the flavor space with the convention as follows: $\Tr(T^{i}T^{j})=\frac{1}{2}\delta^{ij}$ and $T^{0}=\frac{1}{\sqrt{2N_{f}}}\mathbb{1}_{N_{f}\times N_{f}}$ with $N_{f}=2$. Besides the wave function renormalization for the meson, we also introduce one for the quark, i.e., $Z_{q,k}$. Since it plays a minor role in the chiral dynamics in comparison to $Z_{\phi,k}$, the splitting of $Z_{q,k}$ into the transversal and longitudinal components are neglected for simplicity in calculations. Finally, $\hat\mu=\mathrm{diag}(\mu_u,\mu_d)$ in the first line on the r.h.s. of \Eq{eq:action} denotes the matrix of the quark chemical potential, and $\mu=\mu_u=\mu_d$ is assumed in the following unless stated specifically. $A_0$ is the temporal gluon background field, through which the Polyakov dynamics is taken into account.

The RG scale $k$ in \Eq{eq:action} is an infrared cutoff, below which quantum fluctuations are suppressed in the effective action. The evolution of the effective action with $k$ is described by the Wetterich equation \cite{Wetterich:1992yh}, which reads 
\begin{align}
  \partial_t\Gamma_k[\Phi]&=\frac{1}{2}\mathrm{Tr}\big(G_{\phi\phi,k}\partial_t R^{\phi}_{k}\big)-\mathrm{Tr}\big(G_{q\bar{q},k}\partial_t R^{q}_{k}\big)\,, \label{eq:WetterichEqPQM}
\end{align}
with the RG time $t=\ln (k/\Lambda)$ and the initial ultraviolet (UV) cutoff $\Lambda$, where $R^{\phi}_{k}$ and $R^{q}_{k}$ are the regulators for the meson and quark fields, respectively, and they are given in Eqs.~(\ref{eq:Rphi}) and ~(\ref{eq:Rq}). The scale dependent meson and quark propagators are given by
\begin{align}
  G_{\phi\phi/q\bar{q}}[\Phi]=\left( \frac{1}{\frac{\delta^2\Gamma_k[\Phi]}{\delta\Phi^2}+R^{\Phi}_{k}} \right)_{\phi\phi/q\bar{q}}\,, \label{eq:props}
\end{align}
with $\Phi=(q,\bar q,\phi)$ denoting all species of fields.

Inserting the effective action in \Eq{eq:action} into the flow equation \eq{eq:WetterichEqPQM}, one arrives at the flow equation for the effective potential, which reads
\begin{align}
  \partial_t V_k(\rho)=&\frac{k^4}{4\pi^2} \bigg [\big(N^2_f-1\big) l^{(B,4)}_{0}(\bar{m}^{2}_{\pi,k},\eta^{\perp}_{\phi,k},z_\phi;T)\nonumber\\[2ex]
&+l^{(B,4)}_{0}(\bar{m}^{2}_{\sigma,k},\eta^{\perp}_{\phi,k},z_\phi;T)\nonumber\\[2ex]
&-4N_c N_f l^{(F,4)}_{0}(\bar{m}^{2}_{q,k},\eta_{q,k};T,\mu)\bigg]\,, \label{eq:flowV}
\end{align}
with the RG invariant dimensionless meson and quark masses as follows
\begin{align}
  \bar{m}^{2}_{\pi,k}&=\frac{V'_k(\rho)}{k^2Z^{\perp}_{\phi,k}}\,, \qquad \bar{m}^{2}_{\sigma,k}=\frac{V'_k(\rho)+2\rho V''_k(\rho)}{k^2 Z^{\perp}_{\phi,k}}\,,\\[2ex]
  \bar{m}^{2}_{q,k}&=\frac{h^{2}_{k}\rho}{2k^2Z^{2}_{q,k}}\,.
\end{align}
The threshold functions $l^{(B)}_{0}$ and $l^{(F)}_{0}$ are presented in Eqs. (\ref{eq:l0B}) and (\ref{eq:l0F}), and the anomalous dimensions for the mesons and quark are defined as
\begin{align}
  \eta_{\phi,k}^{\perp}&=-\frac{\partial_t Z_{\phi,k}^{\perp}}{Z_{\phi,k}^{\perp}}\,,\quad \eta_{\phi,k}^{\parallel}=-\frac{\partial_t Z_{\phi,k}^{\parallel}}{Z_{\phi,k}^{\parallel}}\,,\quad \eta_{q,k}=-\frac{\partial_t Z_{q,k}}{Z_{q,k}}\,. \label{}
\end{align}
Note that $z_\phi \equiv Z_{\phi,k}^{\parallel}/Z_{\phi,k}^{\perp}$ enters into the flow of $V_k(\rho)$ through the mesonic fluctuations as shown in \Eq{eq:flowV}. 

The transversal anomalous dimension for the  $\pi$-meson is obtained by employing the projection as follows
\begin{align}
  \eta_{\phi,k}^{\perp}&=-\frac{1}{3Z_{\phi,k}^{\perp}}\delta_{ij}\frac{\partial}{\partial (|\bm{p}|^2)}\frac{\delta^2 \partial_t \Gamma_k}{\delta \pi_i(-p) \delta \pi_j(p)}\Bigg|_{\substack{p_0=0\\ \bm{p}=0}}\,,\label{eq:etaphiperp}
\end{align}
and the longitudinal one reads
\begin{align}
  \eta_{\phi,k}^{\parallel}&=-\frac{1}{3Z_{\phi,k}^{\parallel}}\delta_{ij}\frac{\partial}{\partial (p_0^2)}\frac{\delta^2 \partial_t \Gamma_k}{\delta \pi_i(-p) \delta \pi_j(p)}\Bigg|_{\substack{p_0=0\\ \bm{p}=0}}\,.\label{eq:etaphipara}
\end{align}
Their explicit expressions are given in Eqs. (\ref{eq:etaphiperp2}) and (\ref{eq:etaphipara2}), respectively. The difference of the anomalous dimension between the $\pi$ and $\sigma$ mesons is neglected here. Note that even they are different, choosing the $\pi$-meson anomalous dimension for $\eta_\phi$, as done in this work, could minimize the errors of calculation, at least when the baryon chemical potential is not very high, since the mass of the pion is less than that of $\sigma$-meson, because of its nature of Goldstone particle, and the number of the degrees of freedom for the pions is also larger. Therefore, the mesonic degrees of freedom are dominated by the pions. But we should mention that, in the region of  high baryon chemical potential in the phase diagram, especially near the CEP, the $\sigma$-mode is the most relevant collective mode and the mass of the $\sigma$-meson is vanishing, it is necessary to distinguish the anomalous dimensions of $\pi$- and $\sigma$-mesons, which will be investigated in the future.

The anomalous dimension for the quark is obtained by projecting the inverse quark propagator onto the vector channel, as follows
\begin{align}
  \eta_{q}&(p_0,\bm{p})=\frac{1}{4 Z_{q,k}(p_0,\bm{p})}\nonumber \\[2ex]
          &\times\mathrm{Re}\left[\frac{\partial}{\partial (|\bm{p}|^2)}\mathrm{tr}
            \left(i \bm{\gamma}\cdot\bm{p}\left(-\frac{\delta^2}{\delta\bar{q}(p)
            \delta q(p)}\partial_t \Gamma_k\right)\right)\right]\Bigg|_{\substack{p_{0,ex}\\ \bm{p}=0}}\,.   \label{eq:etapsi}
\end{align}
where the spatial component of the external momentum is chosen to be vanishing, as same as the mesonic anomalous dimensions in Eqs. (\ref{eq:etaphiperp}) and (\ref{eq:etaphipara}). Note that because of the fermionic property of the quark, its lowest Matsubara frequency is nonvanishing, and we denote it here as $p_{0,ex}$, which is described in \app{app:anom}.  Projection of the inverse quark propagator onto the scalar channel leads us to the flow of the Yukawa coupling \cite{Pawlowski:2014zaa}, which reads
\begin{align}
  \partial_t h_k(\rho)&=\frac{1}{2 \sigma}\mathrm{Re}\left[\mathrm{tr}\left(-\frac{\delta^2}{\delta\bar{q}(p)
            \delta q(p)}\partial_t \Gamma_k\right)\right]\Bigg|_{\substack{p_{0,ex}\\ \bm{p}=0}}\,.  \label{eq:dth}
\end{align}
The analytic expressions of \Eq{eq:etapsi} and \Eq{eq:dth} are given in \Eq{eq:etapsi2} and \Eq{eq:dth2}, respectively.


\section{Equation of state and baryon number fluctuations}
\label{sec:EoS}

%
\begin{figure*}[t]
\includegraphics[width=1\textwidth]{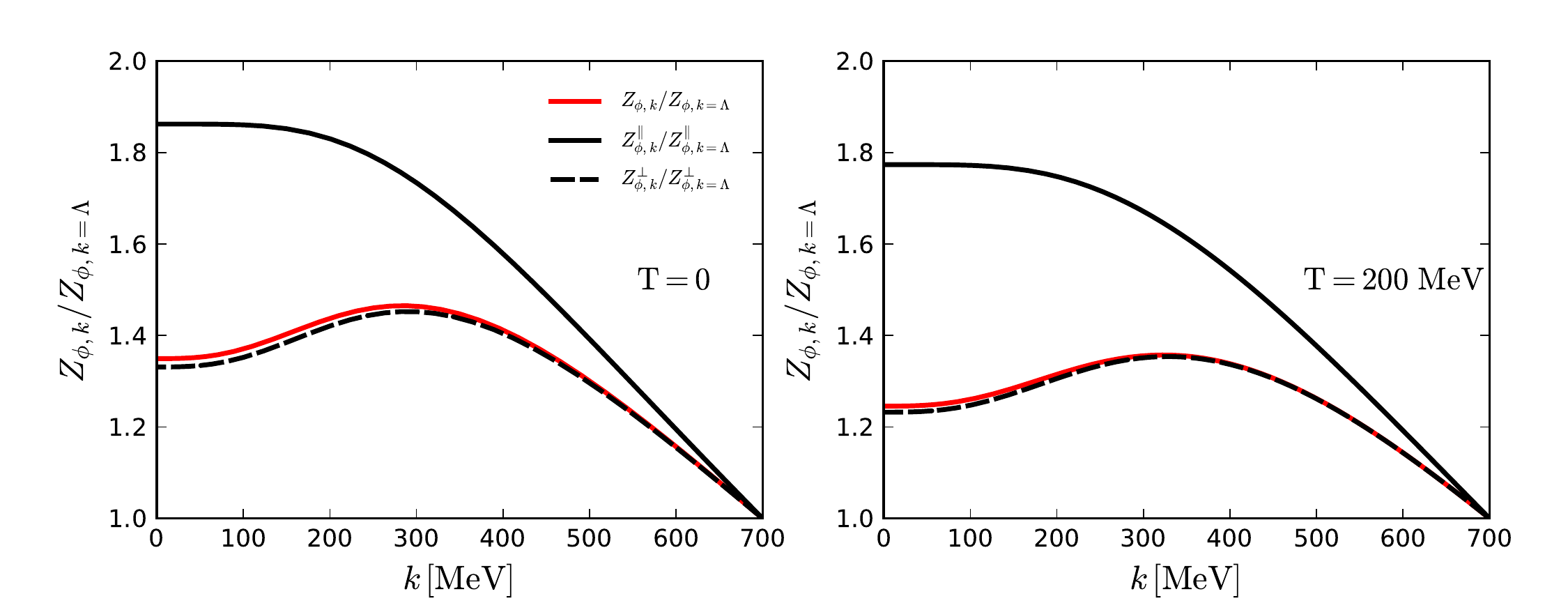}
\caption{Transversal and longitudinal wave function renormalizations for the mesons, $Z_{\phi,k}^{\perp}$, $Z_{\phi,k}^{\parallel}$, as functions of the RG scale $k$ in the vacuum (left panel) and at a finite temperature (right panel), in comparison to the case neglecting the splitting, i.e., $Z_{\phi,k}=Z_{\phi,k}^{\perp}=Z_{\phi,k}^{\parallel}$, as the red line shows. Similar results are also found for finite baryon chemical potentials. Calculations are performed with the fixed point expansion.}\label{fig:zphi}
\end{figure*}
%

%
\begin{figure*}[t]
\includegraphics[width=1\textwidth]{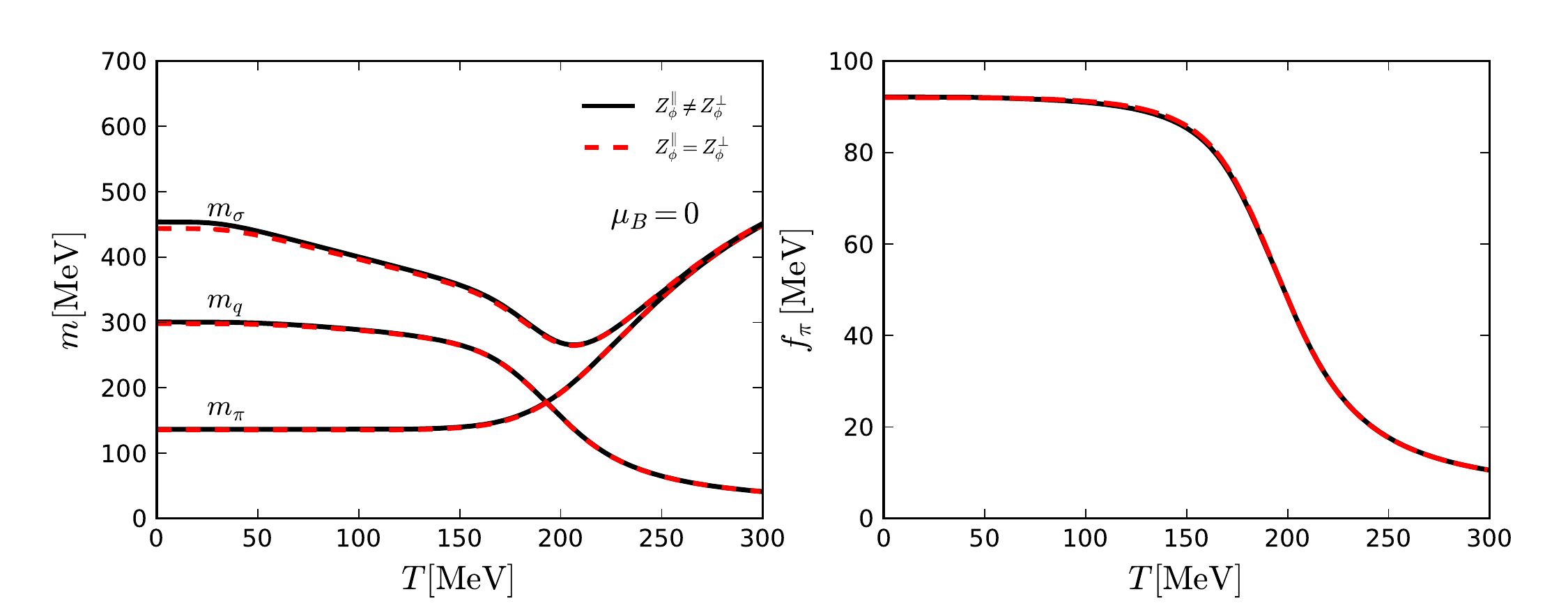}
\caption{Meson and quark masses (left panel) and the $\pi$-meson decay constant $f_\pi$ (right panel) as functions of the temperature at $\mu_B=0$. Calculated results with the splitting of the meson wave function renormalization are compared to those without the splitting, i.e., $Z_{\phi,k}^{\perp}=Z_{\phi,k}^{\parallel}$. Calculations are performed with the fixed point expansion.}\label{fig:mfpi}
\end{figure*}
%

%
\begin{figure}[t]
\includegraphics[width=0.5\textwidth]{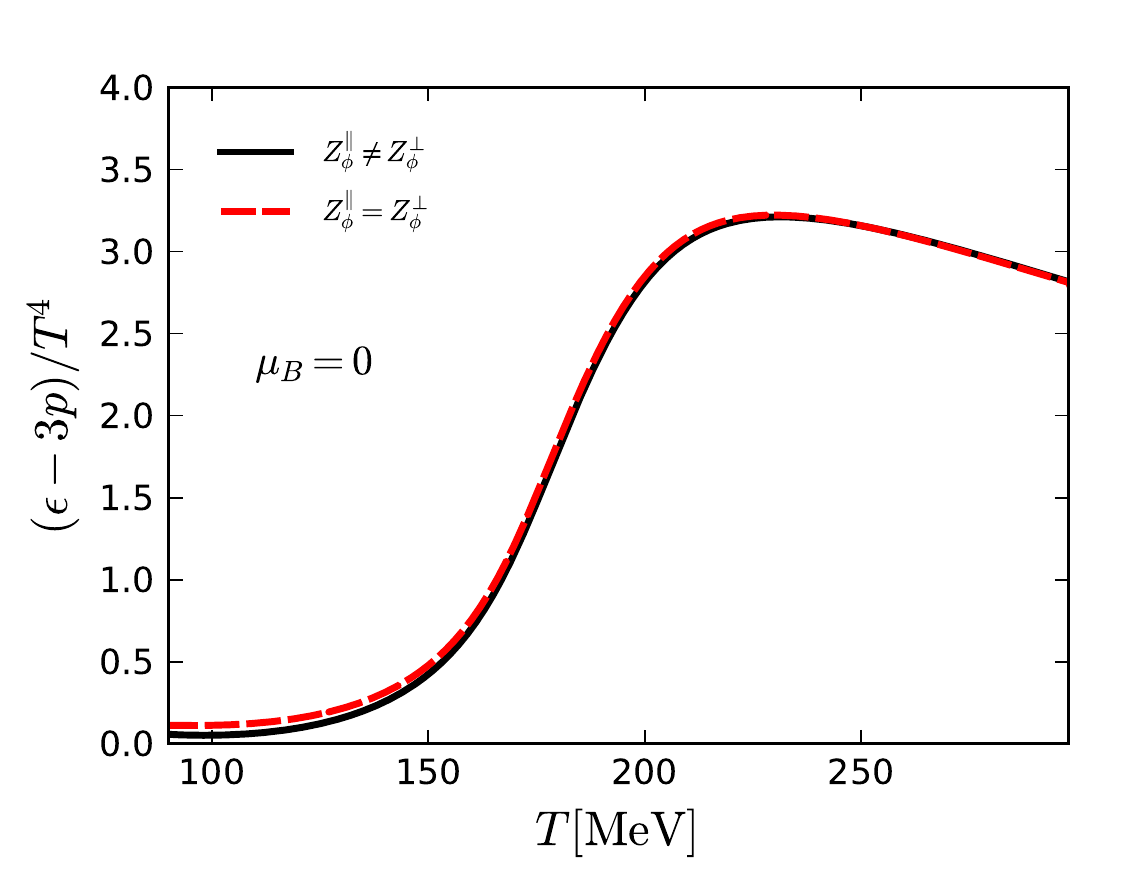}
\caption{Trace anomaly as a function of the temperature for $\mu_B=0$. Calculated results with and without splitting of the wave function renormalization are presented. Calculations are performed with the fixed point expansion.}\label{fig:trace}
\end{figure}
%

%
\begin{figure*}[t]
\includegraphics[width=1\textwidth]{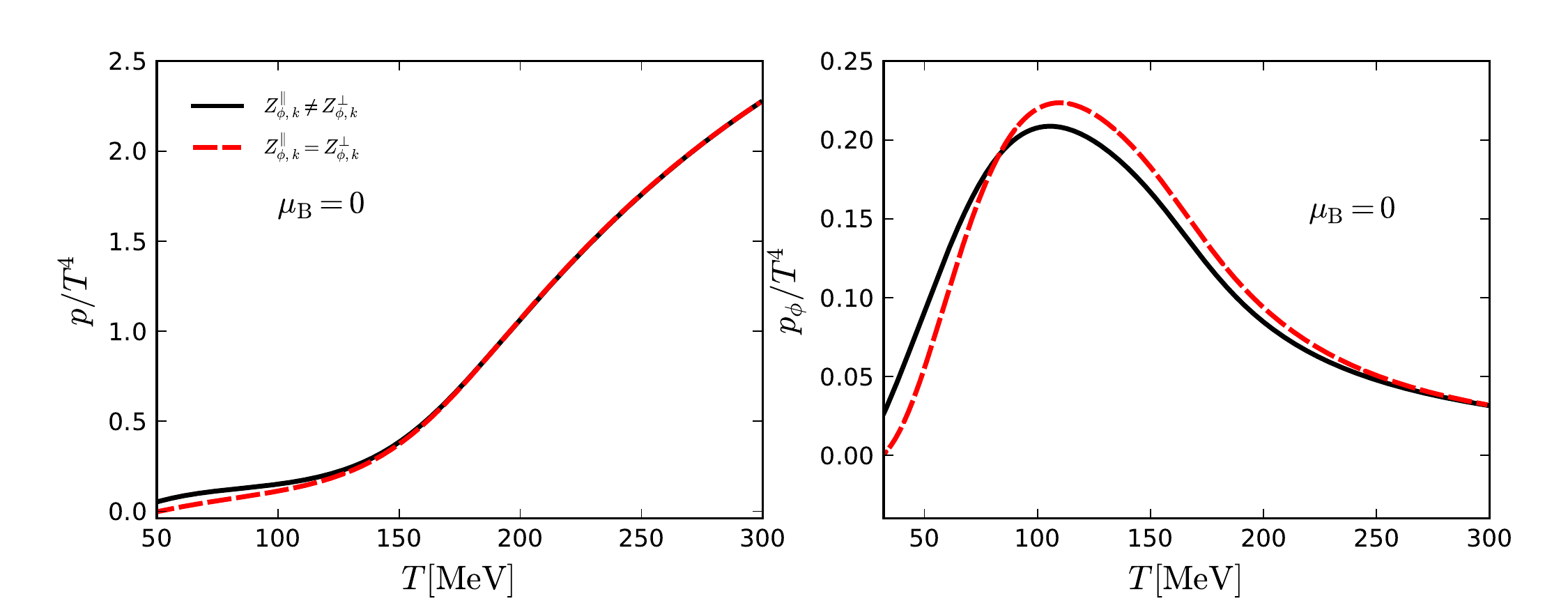}
\caption{Pressure $p$ in \Eq{eq:pres} (left panel) and its contribution from the meson sector $p_\phi$ (right panel), both normalized by $T^4$, as functions of the temperature with $\mu_B=0$. The two cases with and without splitting of the wave function renormalization are compared. Calculations are performed with the fixed point expansion.}\label{fig:pres}
\end{figure*}
%

%
\begin{figure*}[t]
\includegraphics[width=1\textwidth]{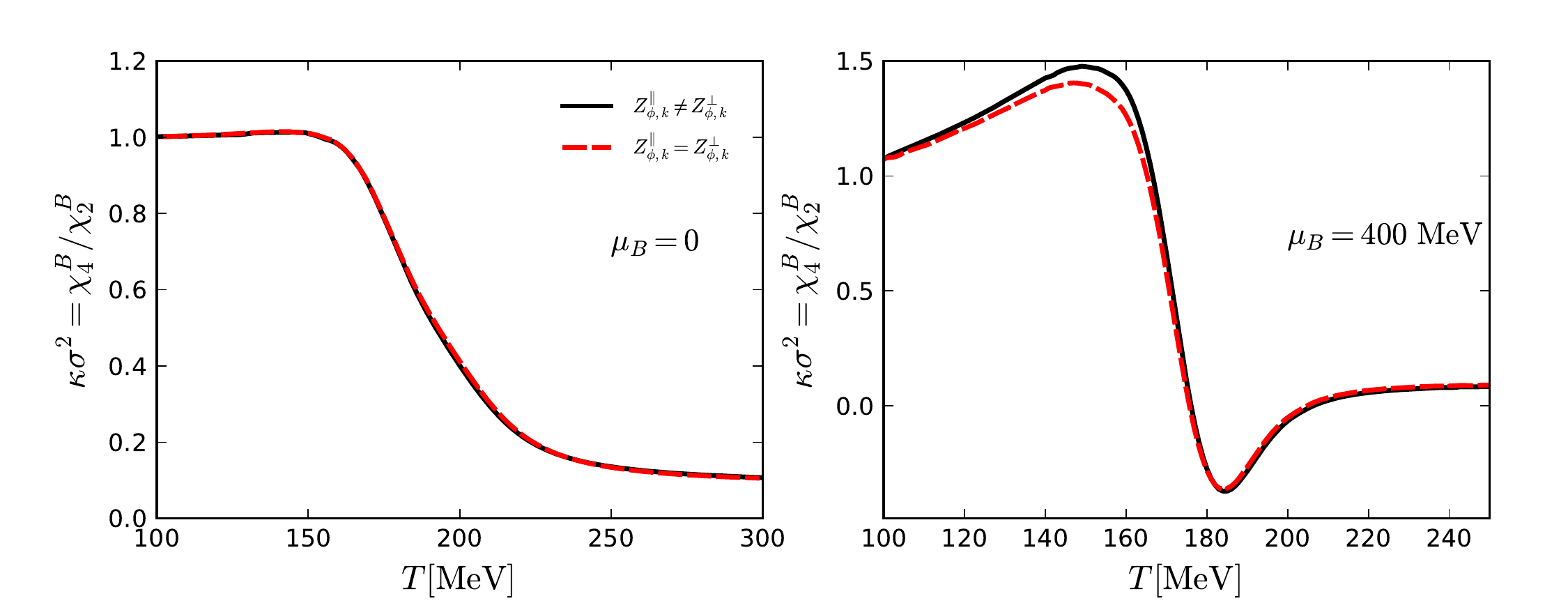}
\caption{Kurtosis of the baryon number distribution $\kappa \sigma^2=\chi_4^{B}/\chi_2^{B}$ as a function of the temperature with $\mu_B=0$ (left panel) and $\mu_B=400$ MeV (right panel).  We compare the results obtained from the splitting of the meson wave function renormalization with those assuming $Z_{\phi,k}^{\perp}=Z_{\phi,k}^{\parallel}$. Calculations are performed with the approach of the fixed point expansion.}\label{fig:kurtosis}
\end{figure*}
%

%
\begin{figure*}[t]
\includegraphics[width=1\textwidth]{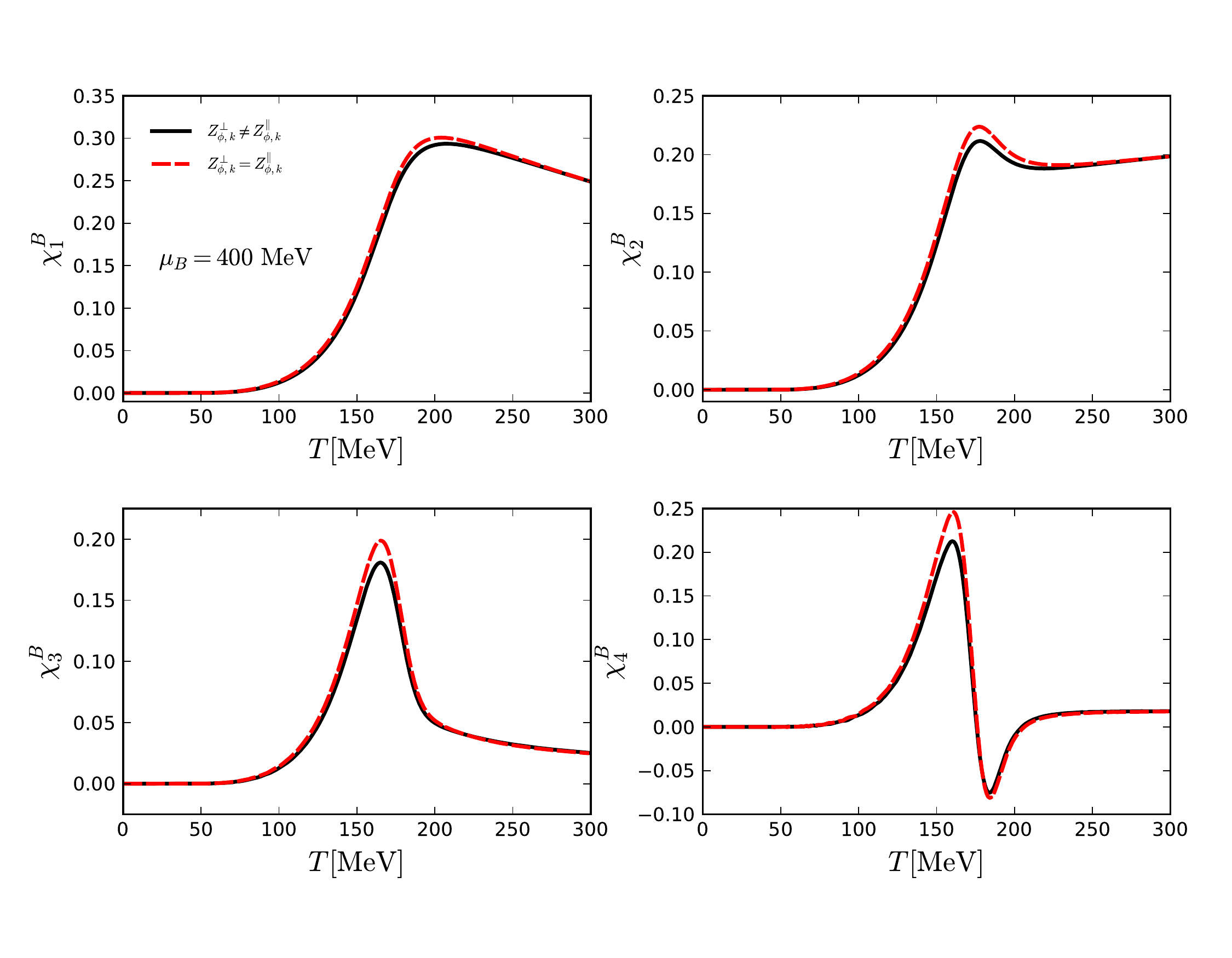}
\caption{Fluctuations of the baryon number, $\chi_n^B$'s, up to the fourth order, as functions of the temperature with  $\mu_B=400$ MeV. In the same way, the two cases, $Z_{\phi,k}^{\perp}\ne Z_{\phi,k}^{\parallel}$ and $Z_{\phi,k}^{\perp}=Z_{\phi,k}^{\parallel}$, are compared. Calculations are performed with the approach of the fixed point expansion.}\label{fig:chi}
\end{figure*}
%

In the PQM model (\ref{eq:action}) the thermodynamical potential density reads
\begin{align}
  \Omega[T,\mu]&=V_{k=0}(\rho)-c\sigma+V_{\text{\tiny{glue}}}(L, \bar L)\,,\label{eq:omega}
\end{align}
where all the fields are on their respective equations of motion, see, e.g., \cite{Fu:2015naa} for more details. $L$ is the traced Polyakov loop and $\bar L$ is its complex conjugate. They are related to the temporal gluonic background field $A_0$ in \Eq{eq:action} through the equations as follows
\begin{align}
L(\bm x)=\frac{1}{N_c}\langle \Tr{\mathcal{P}(\bm x)} \rangle ,\qquad \bar{L} (\bm x)=\frac{1}{N_c}\langle \Tr{\mathcal{P}^{\dagger}(\bm x)} \rangle\,,\label{}
\end{align}
with the Polyakov loop $\mathcal{P}(\bm x)$ which reads
\begin{align}
\mathcal{P}(\bm x)=\mathcal{P}\exp\bigg( ig\int_{0}^{\beta}d\tau A_0(\bm x,\tau) \bigg)\,,\label{}
\end{align}
where the path-ordering operator $\mathcal{P}$ has been employed.

In this work we employ the parametrization of the glue potential in \cite{Lo:2013hla}, which reads 
\begin{align}
  \bar V_{\text{\tiny{glue}}}(L,\bar L)=& -\frac{a(T)}{2} \bar L L + b(T)\ln M_H(L,\bar{L})\nonumber \\[2ex]
  &\quad + \frac{c(T)}{2} (L^3+\bar L^3) + d(T) (\bar{L} L)^2\,,\label{eq:GluepHaar}
\end{align}
with the Haar measure 
\begin{align}
M_H (L, \bar{L})&= 1 -6 \bar{L}L + 4 (L^3+\bar{L}^3) - 3  (\bar{L}L)^2\,,
\end{align}
where $\bar V_{\text{\tiny{glue}}}=V_{\text{\tiny{glue}}}/T^4$ is dimensionless. The temperature dependence of the glue potential is encoded in the coefficients in \Eq{eq:GluepHaar}, which are given by
\begin{align}
  x(T) &= \frac{x_1 + x_2/(t_r+1) + x_3/(t_r+1)^2}{1 + x_4/(t_r+1) + x_5/(t_r+1)^2}\,,\label{eq:xT}
\end{align}
for $x\in \{a, c, d\}$, and 
\begin{align}
  b(T) &=b_1 (t_r+1)^{-b_4}\left (1 -e^{b_2/(t_r+1)^{b_3}} \right)\,,\label{eq:bT}
\end{align}
with the reduced temperature $t_r=(T-T_c)/T_c$. The parameters in the glue potential are determined by fitting the thermodynamics, the Polyakov loop and its fluctuations in the Yang-Mills (YM) theory, and see  \cite{Lo:2013hla} for their values. It has been found that the unquenched effect on the glue potential in QCD is well captured from that in the YM theory \cite{Haas:2013qwp}, by employing the rescale for the reduced temperature as follows
\begin{align}
  (t_r)_{\text{\tiny{YM}}}&\rightarrow \alpha\,(t_r)_{\text{\tiny{glue}}}\,,\label{}
\end{align}
with
\begin{align}
  (t_r)_{\text{\tiny{glue}}}&=(T-T_c^\text{\tiny{glue}})/T_c^\text{\tiny{glue}}\,,\label{}
\end{align}
and $\alpha \simeq 0.57$. $T_c^\text{\tiny{glue}}=250$ MeV is adopted in this work.

The pressure and the energy density are given by
\begin{align}
  p&=-\Omega[T,\mu]\,,\label{eq:pres}\\[2ex]
 \varepsilon&=-p+Ts +\sum_{i=u,d}\mu_i n_i\,,\label{}
\end{align}
with the entropy density and quark number density reading 
\begin{align}
  s&=\frac{\partial p}{\partial T}\qquad \text{and}\qquad n_i=\frac{\partial p}{\partial \mu_i}\,,\label{}
\end{align}
respectively. The interaction measure or the trace anomaly is given as follows
\begin{align}
  \Delta&=\varepsilon-3p\,.\label{}
\end{align}

In this work we will also investigate the fluctuations of the baryon numbers, which are obtained through high-order derivatives of the pressure w.r.t. the baryon chemical potential $\mu_B=3\mu$. The $n$-th order baryon number fluctuation reads
\begin{align}
   \chi_n^{B}&=\frac{\partial^n}{\partial (\mu_B/T)^n}\frac{p}{T^4}\,,\label{eq:suscept}
\end{align}
which is also called as the $n$-th order generalized susceptibility of the baryon number. $ \chi_n^{B}$'s in \Eq{eq:suscept} are related to the cumulants of the baryon number distributions, e.g.,
\begin{align}
  \chi_1^B&=\frac{1}{VT^3}\langle N_B \rangle\,,\\[2ex]
  \chi_2^B&=\frac{1}{VT^3}\langle(\delta N_B)^2\rangle\,,\\[2ex]
  \chi_3^B&=\frac{1}{VT^3}\langle(\delta N_B)^3\rangle\,,\\[2ex]
  \chi_4^B&=\frac{1}{VT^3}\Big(\langle(\delta N_B)^4\rangle-3\langle(\delta N_B)^2\rangle^2\Big)\,,
\end{align}
up to the fourth order. Here $\langle ...\rangle$ denotes the ensemble average and $\delta N_B=N_B-\langle N_B\rangle$. 
In terms of observables in experiments, one has the mean value $M=VT^3\chi_1^{B}$, variance $\sigma^2=VT^3\chi_2^{B}$, skewness $S=\chi_3^{B}/(\chi_2^{B}\sigma)$, and the kurtosis $\kappa=\chi_4^{B}/(\chi_2^{B}\sigma^2)$, etc.


\section{Numerical calculations and results}
\label{sec:num}

%
\begin{figure*}[t]
\includegraphics[width=1\textwidth]{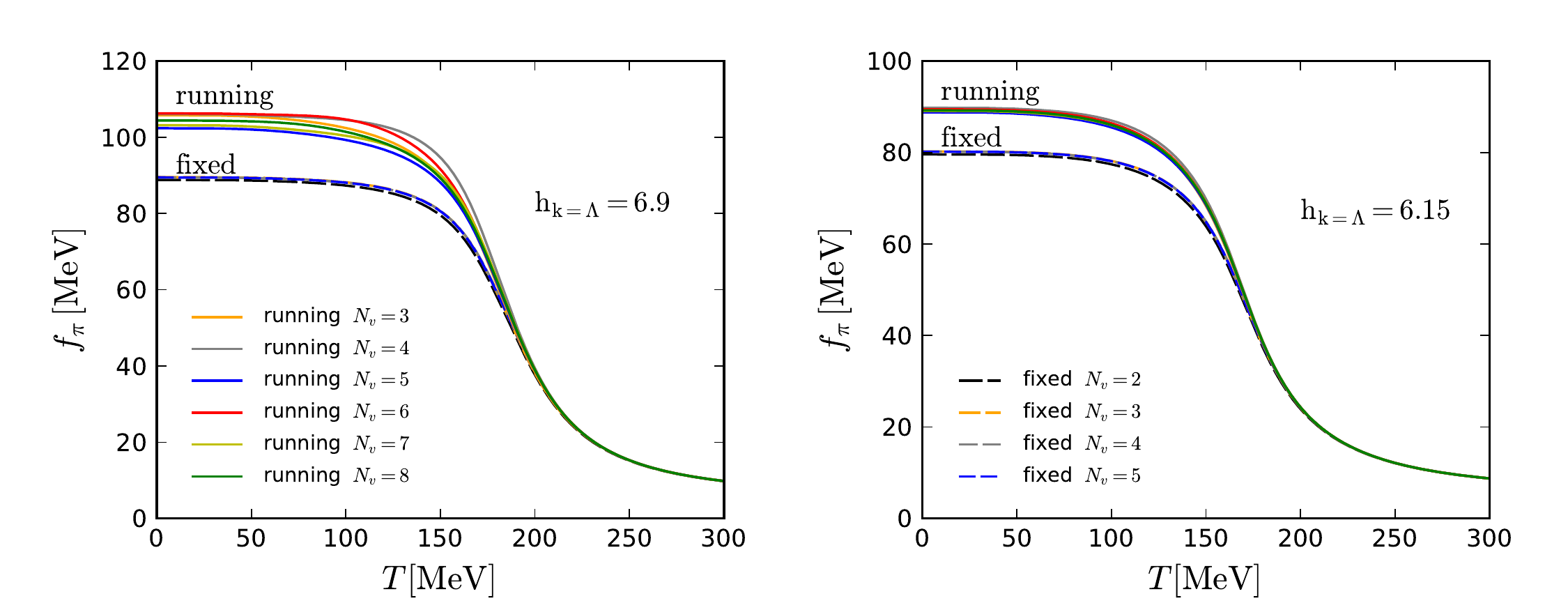}
\caption{$\pi$-meson decay constant $f_\pi$ as a function of the temperature with $\mu_B=0$.  Results obtained from the fixed and running expansion approaches with different values of the maximal order of the Taylor expansion for the effective potential, $N_v$, are presented. $N_h=5$ for the Yukawa coupling is chosen for all the calculations. Two different values of the Yukawa coupling at the UV cutoff, i.e., $h_{k=\Lambda}=$6.90 (left panel) and $h_{k=\Lambda}=$6.15, are adopted in the calculations with other parameters as same as those in \sec{sec:splitting}. Note that we have used the same initial values for the fixed and running expansions.}\label{fig:fpi-expan}
\end{figure*}
%

%
\begin{figure*}[t]
\includegraphics[width=1\textwidth]{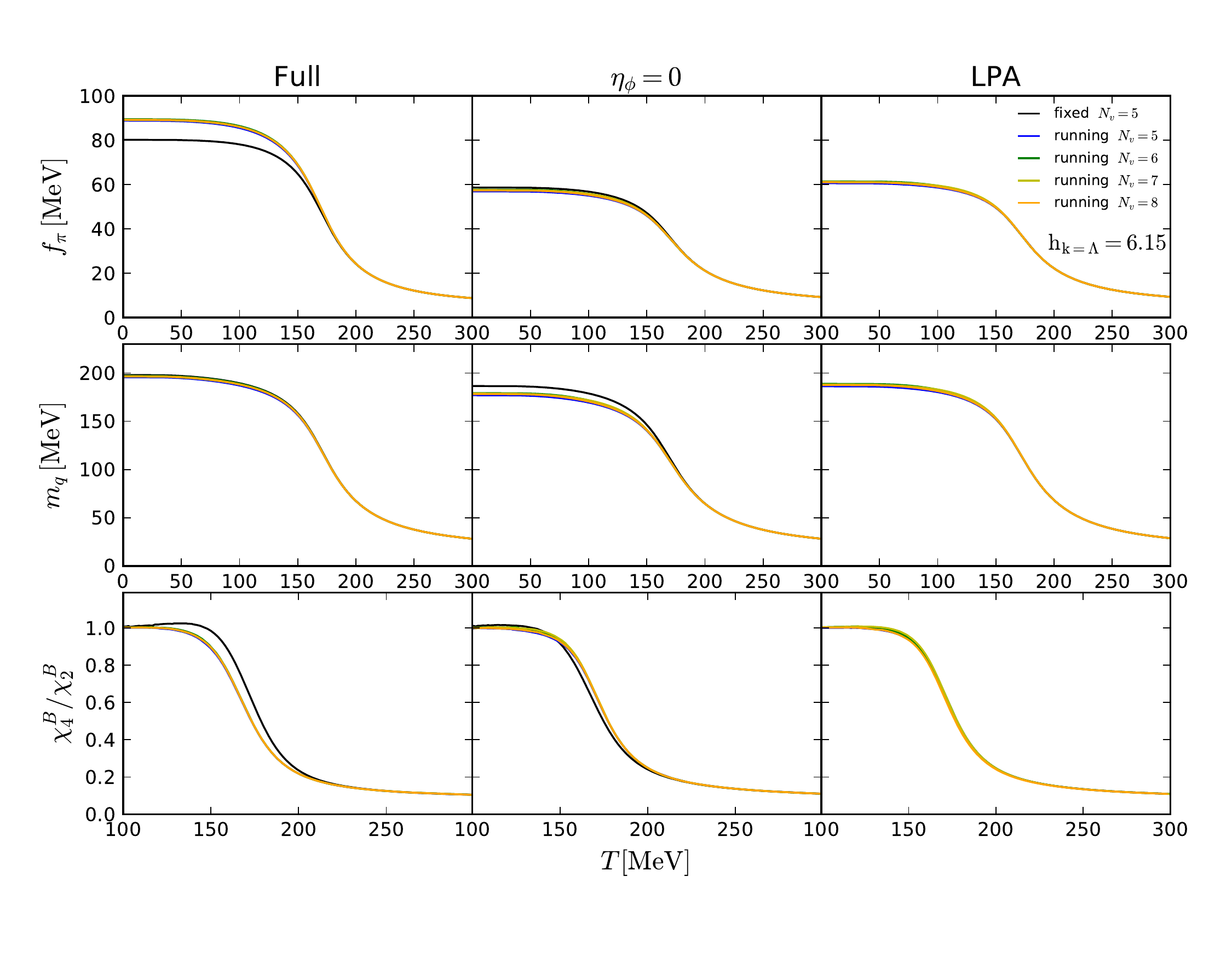}
\caption{Comparison of $f_\pi$, constituent quark mass $m_q$, kurtosis of the baryon number distribution $\chi_4^{B}/\chi_2^{B}$ obtained in the fixed and running expansion approaches as functions of temperature with $\mu_B=0$. The three columns correspond to three different truncations: ``Full'' includes all the ingredients of the truncation described in this paper; ``$\eta_\phi=0$'' denotes the truncation with both the transversal and longitudinal meson anomalous dimensions be vanishing, i.e., $\eta_{\phi,k}^{\perp}=\eta_{\phi,k}^{\parallel}=0$, and others as same as the ``Full'' one; ``LPA'' denotes the standard local potential approximation, where all the anomalous dimensions, including those for the meson and quark, are vanishing, and the Yukawa coupling is just a constant.}\label{fig:trun-expan}
\end{figure*}
%

%
\begin{figure}[t]
\includegraphics[width=0.5\textwidth]{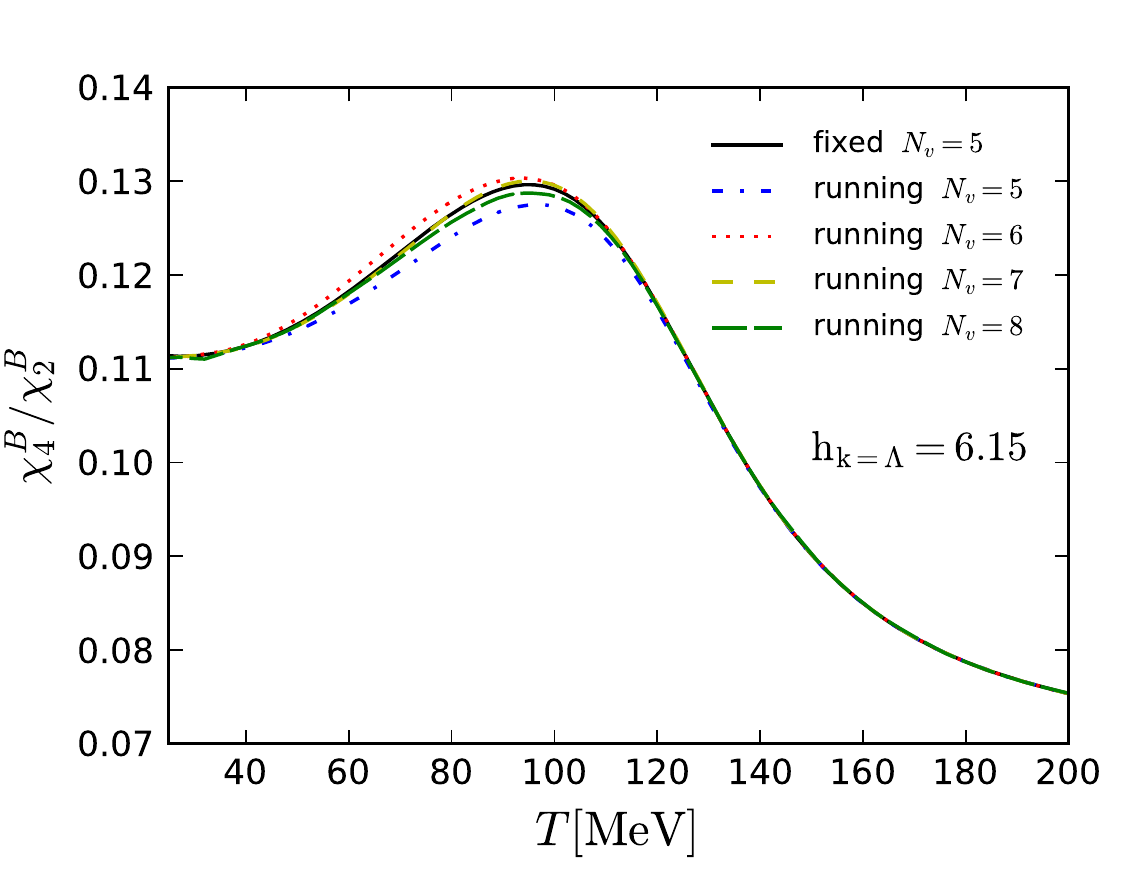}
\caption{Kurtosis of the baryon number distribution $\chi_4^{B}/\chi_2^{B}$ as a function of the temperature with $\mu_B=0$, obtained with $N_v=$5 in the fixed expansion and several values in the running expansion. Calculations are performed in the LPA and with the Polyakov loop $L=\bar L=1$.}\label{fig:R42QM-expan}
\end{figure}
%

We employ the approach of Taylor expansion to solve the flow of the effective potential in \Eq{eq:flowV} numerically. Expanding around the expansion point $\kappa$, one is led to 
\begin{align}\label{}
  V_k(\rho)&=\sum^{N_v}_{n=0}\frac{\lambda_{n,k}}{n!}(\rho-\kappa_k)^n\,,\label{}
\end{align}
where a subscript $k$ is affixed to the expansion point $\kappa$, indicating that the expansion point can be RG scale dependent, and $N_v$ is the highest order which is included in the calculations. The convergency is obtained when the calculated results are almost no longer influenced by the increment of $N_v$, after it is above some value. It is more convenient to work with the renormalized variables, to wit,
\begin{align}
  \bar V_k(\bar \rho)&=\sum^{N_v}_{n=0}\frac{\bar \lambda_{n,k}}{n!}(\bar \rho-\bar \kappa_k)^n\,,\label{eq:barVk}
\end{align}
with $\bar V_k(\bar \rho)=V_k(\rho)$, $\bar \lambda_{n,k}=\lambda_{n,k}/(Z^{\perp}_{\phi,k})^n$, $\bar \rho=Z^{\perp}_{\phi,k} \rho$, and $\bar \kappa_k=Z^{\perp}_{\phi,k}\kappa_k$. Inserting \Eq{eq:barVk} into its flow equation (\ref{eq:flowV}), one is led to
\begin{align}
  &\partial^n_{\bar \rho}\left(\partial_t\big|_{\rho} \bar V_k(\bar \rho)\right)\Big|_{\bar \rho=\bar \kappa_k}\nonumber\\[2ex]
=&(\partial_t -n\eta_{\phi,k}^{\perp})\bar{\lambda}_{n,k}-(\partial_t \bar \kappa_k+\eta_{\phi,k}^{\perp}\bar \kappa_k)\bar \lambda_{n+1,k}\,.\label{eq:drhoV}
\end{align}
Two different expansion approaches are commonly used in the literature: one is the fixed bare expansion point with $\partial_t \kappa_k=0$, i.e., $\kappa_k$ is independent of the scale, which yields 
\begin{align}
  \partial_t \bar \kappa_k+\eta_{\phi,k}^{\perp}\bar \kappa_k&=0\,.\label{eq:dtkappafix}
\end{align}
It follows immediately that the second term on the r.h.s. of \Eq{eq:drhoV} is vanishing for the fixed point expansion. Consequently, $\bar \lambda_{n,k}$'s of different orders are decoupled and there is no feedback from the high order coupling to low order flows. Therefore,  the convergency property is well controlled in the fixed point expansion. For more discussions about the fixed point expansion, see, e.g. \cite{Pawlowski:2014zaa}. Another approach is the running physical expansion, which demands that the expansion point is the solution of equation of motion for every value of $k$, to wit,
\begin{align}
  \frac{\partial}{\partial \bar \rho}\Big(\bar V_k(\bar \rho)-\bar c_k (2\bar \rho)^{\frac{1}{2}}\Big)\bigg|_{\bar \rho=\bar \kappa_k}&=0\,,\label{eq:eosrho}
\end{align}
with $\bar c_k=c/(Z^{\perp}_{\phi,k})^{1/2}$ being the renormalized strength of the explicit chiral symmetry breaking in the effective action \eq{eq:action}. Note that the bare $c$ is scale independent, thus one has $\partial_t \bar c_k=(1/2)\eta_{\phi,k}^{\perp}\bar c_k$. Combining \Eq{eq:drhoV} and \Eq{eq:eosrho}, 
one arrives at the flow of the physical expansion point, which reads
\begin{align}
  \partial_t \bar \kappa_k&=-\frac{\bar c_k^2}{\bar{\lambda}_{1,k}^3+\bar c_k^2\bar{\lambda}_{2,k}}\bigg[\partial_{\bar \rho}\left(\partial_t\big|_{\rho} \bar V_k(\bar \rho)\right)\Big|_{\bar \rho=\bar \kappa_k}\nonumber \\[2ex]
          &+\eta_{\phi,k}^{\perp}\left(\frac{\bar{\lambda}_{1,k}}{2}+\bar\kappa_k\bar{\lambda}_{2,k}\right)\bigg]\,.\label{}
\end{align}
In this work, we will investigate the influence of these two expansion approaches on the fluctuations of the baryon number. 

Furthermore, we also use the Taylor expansion to include the field dependence of the Yukawa coupling, which reads
\begin{align}
  \bar h_k(\bar \rho)&=\sum^{N_h}_{n=0}\frac{\bar h_{n,k}}{n!}(\bar \rho -\bar \kappa_k)^n\,,\label{eq:hTaylor}
\end{align}
with 
\begin{align}
  \bar h_k(\bar \rho)&=\frac{h_k(\rho)}{Z_{q,k}(Z^{\perp}_{\phi,k})^{1/2}}\,,\quad \text{and}\quad \bar h_{n,k}=\frac{h_{n,k}}{Z_{q,k}(Z^{\perp}_{\phi,k})^{(2n+1)/2}}\,,\label{eq: barhk}
\end{align}
where $N_h$ is the order, up to which the Taylor expansion of $\bar h_k(\bar \rho)$ is done. Inserting \Eq{eq: barhk} into \Eq{eq:dth}, one is led to
\begin{align}
  &\partial^n_{\bar \rho}\left(\partial_t\big|_{\rho} \bar h_k(\bar \rho)\right)\Big|_{\bar \rho=\bar \kappa_k}\nonumber\\[2ex]
=&(\partial_t -n\eta_{\phi,k}^{\perp})\bar h_{n,k}-(\partial_t \bar \kappa_k+\eta_{\phi,k}^{\perp}\bar \kappa_k)\bar h_{n+1,k}\,,\label{}
\end{align}
through which the Yukawa couplings of high orders can be solved.

\subsection{Effects of the splitting of the meson wave function renormalization}
\label{sec:splitting}

In this subsection, we investigate the effects of the splitting of meson wave function renormalizations on the equation of state (EoS) and the baryon number fluctuations. In order to focus on this research, only the fixed point expansion for the effective potential in \Eq{eq:dtkappafix} is adopted in this subsection. The maximal orders of the Taylor expansion for the effective potential in \Eq{eq:barVk} and the Yukawa coupling in \Eq{eq:hTaylor} are chosen to be $N_v=5$ and $N_h=5$, respectively. In \sec{sec:expansion} one will see that these values are large enough to obtain convergent results in the fixed expansion, see also \cite{Pawlowski:2014zaa}.

It is left to specify the parameters in the low energy effective models: the UV cutoff is chosen to be $\Lambda=$ 700 MeV; the effective potential at $k=\Lambda$ is parameterized as follows
\begin{align}
  V_\Lambda(\rho)&=\frac{\lambda_\Lambda}{2}\rho^2+\nu_\Lambda\rho\,,\label{}
\end{align}
with $\lambda_\Lambda=$10 and $\nu_\Lambda=(0.556\,\mathrm{GeV})^2$. Furthermore, the strength of the explicit chiral symmetry breaking in \eq{eq:action} is $c=1.97\times 10^{-3}\,(\mathrm{GeV})^3$, and the Yukawa coupling is $h_\Lambda=7.33$. Employing the fixed point expansion and all the truncations discussed in this work, including the splitting of the meson wave function renormalization and the field dependent Yukawa coupling, etc., we obtain the physical observables in the vacuum, which read $f_\pi=$92 MeV, $m_q=$300 MeV, $m_\pi=$136 MeV, and $m_\sigma=$454 MeV. 

In \Fig{fig:zphi} we show the running of the transversal and longitudinal wave function renormalizations for the mesons, $Z_{\phi,k}^{\perp}$, $Z_{\phi,k}^{\parallel}$, with the RG scale $k$. The results obtained from the calculation without splitting are also presented for comparison. Note that values of the wave function renormalizations at the UV cutoff are irrelevant for the computation, since only the anomalous dimensions, rather the wave function renormalizations themselves, enter into the flow equations. One can see that the splitting of $Z_{\phi,k}^{\perp}$ and $Z_{\phi,k}^{\parallel}$ takes place even in the vacuum, which is attributed to $3d$ regulators used in this work, as shown in \Eq{eq:Rphi} and \Eq{eq:Rq}, which lead to partial loss of the $O(4)$ symmetry in the vacuum. We also find that when the splitting is not taken into account, which is usually adopted in the literature, the difference between $Z_{\phi,k}$ and $Z_{\phi,k}^{\perp}$ is very small. Note that $Z_{\phi,k}$ here is also derived from the spatial component, see \Eq{eq:etaphiperp}. Even though, the small difference between $Z_{\phi,k}$ and $Z_{\phi,k}^{\perp}$ has already indicated that the approximation neglecting the splitting is reasonable. In \Fig{fig:mfpi} we investigate the influence of the splitting of the meson wave function renormalization on the meson, quark masses and the $\pi$-meson decay constant $f_\pi$. One observes that there is almost no difference between the two cases with and without splitting, except for a small effect for the $\sigma$-meson mass in the low temperature regime. In \Fig{fig:trace} we plot the trace anomaly as a function of the temperature with $\mu_B=0$. In the same way, we compare the results with and without splitting of the meson wave function renormalization, and the results show that the effect of the splitting on the trace anomaly is small. In the left panel of \Fig{fig:pres}, we investigate the splitting effect of the meson wave function renormalization on the pressure at finite temperature and vanishing baryon chemical potential. Only a mild difference between the two cases with and without splitting is observed in the low temperature regime, which is similar with the result of the trace anomaly. However, we should remind that, in comparison to mesons, the pressure is usually dominated by the degrees of freedom of quarks and gluons, here the gluons realized through the glue potential in the low energy effective theory. Therefore, it is necessary to just single out the contribution from the meson sector to the full pressure, where we denote the meson contribution by $p_\phi$, and investigate the splitting effect on $p_\phi$. The relevant result is presented in the right panel of \Fig{fig:pres}. As expected, the magnitude of $p_\phi$ is quite smaller than that of $p$ in the left panel, and the splitting effect becomes more obvious for $p_\phi$.

In order to investigate whether the partial loss of the $O(4)$ symmetry in the vacuum, arising from the $3d$ regulators employed in our work, results in unphysical results at finite temperature. We have also performed all the calculations with a larger splitting of the wave function renormalization, by increasing the magnitude of longitudinal anomalous dimension $\eta_{\phi,k}^{\parallel}$ artificially. We find that with the increase of the splitting, its influence on the thermodynamical observables grows as well. In another word, this indicates that although the $O(4)$ symmetry is partially lost in the vacuum in our setup due to the use of the $3d$ regulators, this defect is not too harmful for extracting the thermal effects resulting from the nontrivial dispersion relations of mesons. Certainly, an improved regulator will be helpful to clarify this issue eventually.

In \Fig{fig:kurtosis} we investigate the influence of the splitting meson wave function on the kurtosis of the baryon number distribution, i.e., $\kappa \sigma^2=\chi_4^{B}/\chi_2^{B}$. One observes that when the baryon chemical potential is vanishing, the two calculations give almost identical results. With the increase of the chemical potential, e.g., $\mu_B=400$ MeV, as shown in the right panel of \Fig{fig:kurtosis}, the splitting meson wave function has begun to play a role, and there is difference between the two calculations, especially in the region near the phase transition. In order to find where the difference comes from. We present the fluctuations of the baryon number of different orders at $\mu_B=400$ MeV in \Fig{fig:chi}. One can see that the fluctuations of the baryon number of all the orders shown here, has been suppressed a bit by the splitting of the meson wave function renormalization during the chiral phase transition. But the magnitude of the difference is mild.

In a short summary,  in this section we have investigated the influences of the splitting meson wave function renormalization on the phase transition, QCD thermodynamics and EoS, and the fluctuations of the baryon number. We find that the influences are small, except that for the baryon number fluctuations at large baryon chemical potential. It seems that the splitting of the meson wave function renormalization is negligible. However, we should remind that the situation may be different when one calculates, e.g., the spectral function of the meson correlator. Taking the $\pi$-meson for instance, the one particle irreducible two point function reads
\begin{align}
  \Gamma^{(2)}_{\pi \pi}(p_0, \bm{p})&=\frac{\delta^2 \Gamma}{\delta \pi(-p) \delta \pi(p)}= Z_{\phi}^{\parallel} p_0^2+Z_{\phi}^{\perp}\bm{p}^2+m_\pi^2\,.\label{eq:Gam2pipi}
\end{align}
It is apparent that $m_\pi$ in the equation above as well as that in \Fig{fig:mfpi} is just the curvature mass, and the more physical relevant masses are the pole mass, and the screening mass, which are obtained by analytically continuing \Eq{eq:Gam2pipi} from the Euclidean to Minkowski space time, and using the conditions as follow
\begin{align}
  &\Gamma^{(2)}_{\pi \pi}(p_0^2=-m_{\pi,\text{\tiny{pole}}}^2, \bm{p}=0)=0\,, \label{eq:mpole}\\[2ex]
  &\Gamma^{(2)}_{\pi \pi}(p_0^2=0, \bm{p}^2=-m_{\pi,\text{\tiny{screening}}}^2)=0\,. \label{eq:screening}
\end{align}
Based on this na\"ive argument, and inserting \Eq{eq:mpole} and \Eq{eq:screening} into \Eq{eq:Gam2pipi}, one is led to $m_{\pi,\text{\tiny{pole}}}^2=m_\pi^2/Z_{\phi}^{\parallel}$ and $m_{\pi,\text{\tiny{screening}}}^2=m_\pi^2/Z_{\phi}^{\perp}$. One can see that the pole and screening masses are related to the wave function renormalization, and accordingly the relevant spectral function is also affected by the wave function renormalization. Since studies on the influence of the splitting wave function renormalization on the spectral functions are beyond the scope of this paper, we will delay them to the work in the future.

\subsection{Comparison between the fixed and running expansion approaches}
\label{sec:expansion}


In this subsection, we would like to investigate the properties of convergency for the fixed and running expansion approaches. We will vary the value of $N_v$ in \Eq{eq:barVk} to investigate whether the convergency is arrived at for each expansion approach. Moreover, a more important and interesting question is that, when the two expansion approaches adopt the same initial conditions, do they produce the same convergent observables? especially the high order ones, such as non-gaussian cumulants of conserved charges, since any small difference for the two approaches could be amplified in the high order observables. The question proposed here is quite nontrivial, because it is not only related to the self-consistency of the theory, but also one can employ the comparison between the two approaches to investigate the importance of the missing effect in our calculation, such as the field dependence of the wave function renormalization, which is beneficial to the improvement of truncation. 

In \Fig{fig:fpi-expan} we have performed the convergency test for both the fixed and running expansions for the effective potential. We find that the convergency for the fixed expansion is very well, and calculations with $N_v\geq 3$ have already produces reliable results. This behavior is also observed in \cite{Pawlowski:2014zaa}. This excellent convergency property for the fixed point expansion is easily understood, because there is no feedback from the high order expansion coefficients to the lower ones due to \Eq{eq:dtkappafix}. The situation for the running expansion is more complicated, and we find that the numerical instability arising from the the feedback is harmful to the convergency, as shown in the left panel of \Fig{fig:fpi-expan}, even the maximal expansion order $N_v$ grows up to eighth, oscillations are still observed there. However, the convergency for the running expansion could be improved significantly, if the system is away from the regime of the numerical instability, as we show in the right panel of \Fig{fig:fpi-expan}, where the smaller initial Yukawa coupling suppresses the numerical instability remarkably, and a good convergency is observed for the running expansion approach. 

In \Fig{fig:trun-expan} we show the $\pi$-meson decay constant $f_\pi$, constituent quark mass $m_q$, and the kurtosis of the baryon number distribution $\chi_4^{B}/\chi_2^{B}$ calculated in the fixed and running expansion approaches. Results obtained with $N_v=5$ for the fixed expansion, and those with $N_v=$5 through 8 for the running expansion are presented for comparison. Note that $N_v=5$ is large enough to obtain the convergency for the fixed expansion, see \Fig{fig:fpi-expan}. In the first column of \Fig{fig:trun-expan} we present the results obtained with the ``Full'' truncation in which all the ingredients in \Eq{eq:action} are encoded.  One observes that the convergency for the running expansion has already been arrived at with $N_v\geq 5$, but the discrepancy between the fixed and running expansion is obvious, in particular for $f_\pi$ and $\chi_4^{B}/\chi_2^{B}$. We have included the field dependence for the effective potential in \Eq{eq:barVk} and the Yukawa coupling in \Eq{eq:hTaylor} in our calculations, but not for the wave function renormalizations or the anomalous dimensions. Therefore, it is reasonable to attribute the discrepancy to the missing field-dependence of the mesonic and quark wave function renormalizations in our computation. In order to make this point more manifest, we turn off the mesonic anomalous dimension, and the relevant results are given in the second column of \Fig{fig:trun-expan}. As we expect, the difference between the fixed and running expansion becomes smaller, especially for $f_\pi$ and $\chi_4^{B}/\chi_2^{B}$. The difference for the quark mass increases a bit, which indicates that the field dependence of the quark anomalous dimension also play a relatively small role there. In the third column of \Fig{fig:trun-expan}, we turn off all the anomalous dimensions and the flow of Yukawa coupling, and employ the standard local potential approximation (LPA). One can see that results from the fixed and running expansion approaches are in good agreement with each other. In \Fig{fig:R42QM-expan} we separate the influence of the gluon dynamics, set the Polyakov loop $L=\bar L=1$, and compare $\chi_4^{B}/\chi_2^{B}$ obtained from the fixed and running expansions. In the same way we adopt the LPA, and one can see the two expansions give consistent results.

\section{Summary and outlook}
\label{sec:sum}

In this work we have studied the nontrivial dispersion relation of mesons resulting from the splitting of the transversal and longitudinal mesonic wave function renormalizations, and its influences on the QCD phase transition, equation of state, the fluctuations of baryon number, etc. The calculations are performed in a two-flavor low energy effective model within the functional renormalization group approach. 

It is found that the influences of the splitting mesonic wave function renormalizations on the QCD phase transition, QCD EoS, and the fluctuations of baryon number are small, except that the fluctuations of the baryon number at high baryon chemical potential are suppressed a bit during the chiral phase transition by the splitting. Note that although the splitting effect on the equilibrium thermodynamical bulk properties is mild, it would potentially play a remarkable role for other observables related to the nonequilibrium dynamics, such as the spectral function of correlators, as discussed at the end of \sec{sec:splitting}, which will be investigated in our future work.

Furthermore, we have investigated the property of convergency for the fixed and running expansion for the effective potential. Since there is no feedback from the high order expansion coefficients to the lower ones, the convergency for the fixed point expansion is very well. On the contrary, the convergency could be spoiled, if there is numerical instability in the calculation by employing the running expansion for the effective potential, but the convergency can be improved significantly, once the calculation is performed far away from the regime of numerical instability. 

Through the comparison between the results obtained from the fixed expansion and those from the running expansion, we find that the field dependence of the meson and quark wave function renormalizations, in particular the former, are important for the quantitative computations, which will be done in our future work.

\begin{acknowledgments}

We thank Jan M. Pawlowski for valuable discussions. The work was supported by the National Natural Science Foundation of China under Contracts Nos. 11775041.

\end{acknowledgments}


\appendix

\section{Anomalous dimensions and Yukawa coupling}
\label{app:anom}

Inserting the effective action in \Eq{eq:action} and its the flow equation in \Eq{eq:WetterichEqPQM} into \Eq{eq:etaphiperp}, and employing the $3d$- regulators in \Eq{eq:Rphi} and \Eq{eq:Rq}, one obtains the transversal anomalous dimension for the meson, as follows
\begin{align}
  \eta_{\phi,k}^{\perp}&=\frac{1}{6\pi^2}\Bigg\{\frac{4}{k^2 z_\phi^4} \bar{\kappa}_k(\bar{V}''_k(\bar{\kappa}_k))^2\mathcal{BB}_{(2,2)}(\bar{m}^{2}_{\pi,k},\bar{m}^{2}_{\sigma,k};T)\nonumber\\[2ex]
&+N_c\bar{h}^{2}_{k}\bigg[\mathcal{F}_{(2)}(\bar{m}^{2}_{q,k};T,\mu)(2\eta_{q,k}-3)\nonumber\\[2ex]
&-4(\eta_{q,k}-2)\mathcal{F}_{(3)}(\bar{m}^2_{q,k};T,\mu)\bigg]\Bigg\}\,, \label{eq:etaphiperp2}  
\end{align} 
and relevant threshold functions are given in \app{app:thresholdfun}. In the same way, the longitudinal anomalous dimension for the meson in \Eq{eq:etaphipara} reads
\begin{align}
  \eta_{\phi,k}^{\parallel}&=\frac{1}{6\pi^2}\Bigg\{\frac{4}{k^2 z_\phi^4}\bar{\kappa}_k(\bar{V}''_k(\bar{\kappa}_k))^2\bigg[-6\mathcal{BB}_{(2,2)}(\bar{m}^{2}_{\pi,k},\bar{m}^{2}_{\sigma,k};T)\nonumber\\[2ex]
&+\frac{4}{z_\phi}(1+\bar{m}^{2}_{\sigma}){\mathcal{BB}}_{(2,3)}(\bar{m}^{2}_{\pi,k},\bar{m}^{2}_{\sigma,k};T)\nonumber\\[2ex]
&+\frac{4}{z_\phi}(1+\bar{m}^{2}_{\pi}){\mathcal{BB}}_{(3,2)}(\bar{m}^{2}_{\pi,k},\bar{m}^{2}_{\sigma,k};T)\bigg]\nonumber\\[2ex]
&\times(1-\frac{1}{5}\eta^{\bot}_{\phi,k})+\frac{N_c\bar{h}^{2}_{k}}{z_\phi}\mathcal{F}_{(3)}(\bar{m}^{2}_{q,k};T,\mu)(4-\eta_{q,k})\bigg\}\,, \label{eq:etaphipara2}  
\end{align} 

The anomalous dimension for the quark in \Eq{eq:etapsi} is given by
\begin{align}
\eta_{q,k}=&\frac{1}{24\pi^2N_f}(4-\eta_{\phi,k}^{\perp})\bar{h}^{2}_{k}\nonumber\\[2ex]
&\times\bigg\{ (N^{2}_{f}-1)\mathcal{FB}_{(1,2)}(\bar{m}^{2}_{q,k},\bar{m}^{2}_{\pi,k};T,\mu,p_{0,ex})\nonumber\\[2ex]
&+\mathcal{FB}_{(1,2)}(\bar{m}^{2}_{q,k},\bar{m}^{2}_{\sigma,k};T,\mu,p_{0,ex})\bigg\}\,, \label{eq:etapsi2}
\end{align} 
where $p_{0,ex}=\pi T$ for the finite temperature part and $\big[k^2+(\pi T)^2\exp\{-2k/(5 T)\}\big]^{1/2}$ for the vacuum part in the threshold function $\mathcal{FB}$'s in \Eq{eq:FB}. Note that the modification of the lowest-order Matsubara frequency in the vacuum part is employed to suppress the artificial temperature dependence of thermodynamics in the low temperature regime, see e.g. \cite{Fu:2015naa,Fu:2016tey} for more discussions. The flow of the field-dependent Yukawa coupling in \Eq{eq:dth} is firstly derived in \cite{Pawlowski:2014zaa} and is also presented here for convenience, which reads
\begin{align}
  \partial_t&\big|_{\rho}\bar{h}_k(\bar{\rho})=\left(\frac{1}{2}\eta_{\phi,k}+\eta_{q,k}\right)\bar{h}_k(\bar{\rho})\nonumber\\[2ex]
&+\frac{\bar{h}^3_k(\bar{\rho})}{4\pi^2N_f}\bigg[L^{(4)}_{(1,1)}(\bar{m}^{2}_{q,k},\bar{m}^{2}_{\sigma,k},\eta_{q,k},\eta_{\phi,k}^{\perp};T,\mu,p_{0,ex})\nonumber\\[2ex]
&-(N^{2}_{f}-1)L^{(4)}_{(1,1)}(\bar{m}^{2}_{q,k},\bar{m}^{2}_{\pi,k},\eta_{q,k},\eta_{\phi,k}^{\perp};T,\mu,p_{0,ex})\bigg]\nonumber\\[2ex]
&+\frac{1}{2\pi^2}\bar{h}_k(\bar{\rho})\bar{h}'_k(\bar{\rho})\bar{\rho}\bigg[\bar{h}_k(\bar{\rho})+\bar{\rho}\bar{h}'_k(\bar{\rho})\bigg]\nonumber\\[2ex]
&\times L^{(4)}_{(1,1)}(\bar{m}^{2}_{q,k},\bar{m}^{2}_{\sigma,k},\eta_{q,k},\eta_{\phi,k}^{\perp};T,\mu,p_{0,ex})\nonumber\\[2ex]
&-\frac{k^2}{4\pi^2}\bigg[\left(3\bar{h}'_k(\bar{\rho})+2\bar{\rho}\bar{h}''_k(\bar{\rho})\right)l^{(B,4)}_{1}(\bar{m}^{2}_{\sigma,k},\eta_{\phi,k}^{\perp};T)\nonumber\\[2ex]
&+3\bar{h}'_k(\bar{\rho})l^{(B,4)}_{1}(\bar{m}^{2}_{\pi,k},\eta_{\phi,k}^{\perp};T)\bigg]\,.\label{eq:dth2}  
\end{align}

\section{Threshold functions}
\label{app:thresholdfun}

In this work we use the $3d$- flat or Litim regulators \cite{Litim:2000ci,Litim:2001up}, which are very suited for the computations at finite temperature and densities, since the summation for the Matsubara frequencies is not affected by the $3d$ regulators, and can be performed analytically. The regulators in \Eq{eq:WetterichEqPQM} read
\begin{align}
  R^{\phi}_{k}(q_0,\bm{q})&=Z^{\perp}_{\phi,k}\bm{q}^2 r_B(\bm{q}^2/k^2)\,, \label{eq:Rphi}\\[2ex] 
  R^{q}_{k}(q_0,\bm{q})&=Z_{q,k}i\bm{\gamma} \cdot \bm{q} r_F(\bm{q}^2/k^2)\,, \label{eq:Rq}
\end{align} 
with 
\begin{align}
  r_B(x)&=\left( \frac{1}{x}-1 \right)\Theta(1-x)\,,\\[2ex] 
  r_F(x)&=\left( \frac{1}{\sqrt{x}}-1 \right)\Theta(1-x)\,,  \label{}
\end{align} 
where $\Theta(x)$ is the Heaviside step function. Note that since $Z^{\perp}_{\phi,k}\neq Z^{\parallel}_{\phi,k}$, it is  the transversal wave function renormalization for the meson appearing in \Eq{eq:Rphi}.

In the threshold functions we usually need the dimensionless meson and quark propagators as follows
\begin{align}
  G_\phi(q,\bar{m}^{2}_{\phi,k})&=\frac{1}{z_\phi\tilde{q}^{2}_{0}+1+\bar{m}^{2}_{\phi,k}}\,, \label{eq:Gphi}\\[2ex] 
  G_q(q,\bar{m}^{2}_{q,k})&=\frac{1}{(\tilde{q}_0+i\tilde{\mu})^2+1+\bar{m}^{2}_{q,k}}\,, \label{eq:Gq}
\end{align} 
with $\tilde{\mu}=\mu/k$ and $\tilde{q}_0=q_0/k$, where the Matsubara frequency is  $q_0=2n_q\pi  T$ for bosons and $(2n_q+1)\pi T $ for fermions with $n_q\in \mathbb{Z}$.

The definition of the threshold functions $\mathcal{B}_{(n)}$ and $\mathcal{F}_{(n)}$ is given by
\begin{align}
  \mathcal{B}_{(n)}(\bar{m}^{2}_{\phi,k};T)&=\frac{T}{k}\sum_{n_q}\Big(G_{\phi}(q,\bar{m}^{2}_{\phi,k})\Big)^n\,,\label{eq:Bn} \\[2ex]
  \mathcal{F}_{(n)}(\bar{m}^{2}_{q,k};T,\mu)&=\frac{T}{k}\sum_{n_q}\Big(G_q(q,\bar{m}^{2}_{q,k})\Big)^n\,.\label{eq:Fn}
\end{align}
After performing the Matsubara sum, one arrives at
\begin{align}
    \mathcal{B}_{(1)}(\bar{m}^{2}_{\phi,k};T)=&\frac{1}{\sqrt{z_\phi(1+\bar{m}^{2}_{\phi,k})}}\bigg(\frac{1}{2}+n_{B}(\bar{m}^{2}_{\phi,k},z_\phi;T)\bigg)\,,\label{eq:B1}
\end{align}
and
\begin{align}
    &\mathcal{F}_{(1)}(\bar{m}^{2}_{q,k};T,\mu)=\frac{1}{2\sqrt{1+\bar{m}^{2}_{q,k}}}\nonumber\\[2ex]
    &\hspace{.3cm}\times\Big(1-n_{F}(\bar{m}^{2}_{q,k};T,\mu)-n_{F}(\bar{m}^{2}_{q,k};T,-\mu)\Big)\,,\label{eq:F1}
\end{align}
with the bosonic and fermionic distribution functions being
\begin{align}
  n_B(\bar{m}^{2}_{\phi,k},z_\phi;T)&=\frac{1}{\exp\bigg\{\frac{1}{T}\frac{k}{z_\phi^{1/2}}\sqrt{1+\bar{m}^{2}_{\phi,k}}\bigg\}-1}\,,
\end{align} 
and
\begin{align}
  n_F(\bar{m}^{2}_{q,k};T,\mu)&=\frac{1}{\exp\bigg\{\frac{1}{T}\Big[k\sqrt{1+\bar{m}^{2}_{q,k}}-\mu\Big]\bigg\}+1}\,.
\end{align} 
Note that when the Polaykov loop is taken into account, the fermionic distribution function is modified as follows
\begin{align}
n_F(\bar{m}^{2}_{q,k};T,\mu,L,\bar{L})&=\frac{1+2\bar{L}e^{x/T}+Le^{2x/T}}{1+3\bar{L}e^{x/T}+3Le^{2x/T}+e^{3x/T}}\,,
\end{align} 
with $x=k\sqrt{1+\bar{m}^{2}_{q,k}}-\mu$, and $n_{F}(\bar{m}^{2}_{q,k};T,-\mu)$ in \Eq{eq:F1} is replaced with $n_F(\bar{m}^{2}_{q,k};T,-\mu,\bar L, L)$ accordingly. With \Eq{eq:B1} and \Eq{eq:F1}, high-order threshold functions in Eqs. (\ref{eq:Bn}) and (\ref{eq:Fn}) are readily obtained as
\begin{align}
  \mathcal{B}_{(n+1)}(\bar{m}^{2}_{\phi,k};T)&=-\frac{1}{n}\frac{\partial}{\partial \bar{m}^{2}_{\phi,k}}\mathcal{B}_{(n)}(\bar{m}^{2}_{\phi,k};T)\,,\\[2ex]
  \mathcal{F}_{(n+1)}(\bar{m}^{2}_{q,k};T,\mu)&=-\frac{1}{n}\frac{\partial}{\partial \bar{m}^{2}_{q,k}}
                                    \mathcal{F}_{(n)}(\bar{m}^{2}_{q,k};T,\mu)\,. 
\end{align}

The threshold functions in the flow of effective potential in \Eq{eq:flowV} are given by
\begin{align}
  &l_0^{(B,d)}(\bar{m}^{2}_{\phi,k},\eta^\perp_{\phi,k},z_\phi;T)\nonumber\\[2ex]
  =&\frac{2}{d-1}\left( 1- \frac{\eta^{\bot}_{\phi,k}}{d+1}\right)\mathcal{B}_{(1)}(\bar{m}^{2}_{\phi,k},z_\phi;T)\,,\label{eq:l0B}
\end{align} 
and
\begin{align}
  &l_0^{(F,d)}(\bar{m}^{2}_{q,k},\eta_{q,k};T,\mu)\nonumber\\[2ex]
  =&\frac{2}{d-1}\left( 1-\frac{\eta_{q,k}}{d} \right)\mathcal{F}_{(1)}(\bar{m}^{2}_{q,k};T,\mu)\,.\label{eq:l0F}
\end{align} 
In \Eq{eq:dth2}, one also needs $l^{(B,4)}_{1}$, which is obtained from 
\begin{align}
  l_1^{(B/F,d)}(m^{2})=&-\frac{\partial}{\partial m^{2}}l_0^{(B/F,d)}(m^{2})\,.\label{}
\end{align} 

Furthermore, we also need other threshold functions, such as
\begin{align}
  &\mathcal{BB}_{(n_1,n_2)}(m_1^2,m_2^2;T)\nonumber\\[2ex]
=&\frac{T}{k}\sum_{n_q}\Big(G_\phi(q,\bar{m}^{2}_{\phi_a,k})\Big)^{n_1} \Big(G_\phi(q,\bar{m}^{2}_{\phi_b,k})\Big)^{n_2}\,.\label{eq:BBn1n2}
\end{align}
in the expressions of the mesonic anomalous dimension in Eqs. (\ref{eq:etaphiperp}) and (\ref{eq:etaphipara}). Inserting \Eq{eq:Gphi} into \Eq{eq:BBn1n2}, one is led to
\begin{align}
  &\mathcal{BB}_{(1,1)}(\bar{m}^{2}_{\phi_a,k},\bar{m}^{2}_{\phi_b,k};T)\nonumber\\[2ex]
  =&-\frac{1}{z^{1/2}_{\phi}}\Bigg\{\left(\frac{1}{2}+n_B(\bar{m}^{2}_{\phi_a,k},z_\phi;T) \right)\frac{1}{(1+\bar{m}^{2}_{\phi_a,k})^{1/2}} \nonumber\\[2ex]
&\times \frac{1}{\bar{m}^{2}_{\phi_a,k}-\bar{m}^{2}_{\phi_b,k}}+\left( \frac{1}{2}+n_B(\bar{m}^{2}_{\phi_b,k},z_\phi;T) 
\right)\nonumber\\[2ex]
&\times \frac{1}{(1+\bar{m}^{2}_{\phi_b,k})^{1/2}}\frac{1}{\bar{m}^{2}_{\phi_b,k}-\bar{m}^{2}_{\phi_a,k}}\Bigg\}\,.\label{eq:BB11}
\end{align} 
In the same way, one could obtain higher-order ones by employing, e.g.,
\begin{align}
  &\mathcal{BB}_{(n_1+1,n_2)}(\bar{m}^{2}_{\phi_a,k},\bar{m}^{2}_{\phi_b,k};T)\nonumber\\[2ex]
  =&-\frac{1}{n_1}\frac{\partial}{\partial \bar{m}^{2}_{\phi_a,k}}\mathcal{BB}_{(n_1,n_2)}(\bar{m}^{2}_{\phi_a,k},\bar{m}^{2}_{\phi_b,k};T)\,.
\end{align}

In the flow of the Yukawa coupling in (\ref{eq:dth}), the threshold function $L$ is introduced, which reads
\begin{align}
  &L^{(4)}_{(1,1)}(\bar{m}^{2}_{q,k},\bar{m}^{2}_{\phi,k},\eta_{q,k},\eta_{\phi,k};T,\mu,p_0)\nonumber\\[2ex]
  =&\frac{2}{3}\bigg[\left(1-\frac{\eta_{\phi,k}^{\perp}}{5}\right)\mathcal{FB}_{(1,2)}(\bar{m}^{2}_{q,k},\bar{m}^{2}_{\phi,k};T,\mu,p_0)\nonumber\\[2ex]
&+\left(1-\frac{\eta_{q,k}}{4}\right)\mathcal{FB}_{(2,1)}(\bar{m}^{2}_{q,k},\bar{m}^{2}_{\phi,k};T,\mu,p_0)\bigg]\,,
\end{align} 
where the fermionic and bosonic mixing threshold functions $\mathcal{FB}$'s are given by
\begin{align}
 &\mathcal{FB}_{(n_f,n_b)}(\bar{m}^{2}_{q,k},\bar{m}^{2}_{\phi,k};T,\mu,p_0)\nonumber \\[2ex] 
  =&\frac{T}{k}\sum_{n_q}\Big(G_{q}(q,\bar{m}^{2}_{q,k})\Big)^{n_f}\Big(G_{\phi}(q-p,\bar{m}^{2}_{\phi,k})\Big)^{n_b}\,.\label{eq:FB}
\end{align}
The explicit expression for $\mathcal{FB}_{(1,1)}$ is readily obtained after summing the Matsubara frequency, which can be found in, e.g. \cite{Fu:2015naa}. Higher-order $\mathcal{FB}$'s are obtained from $\mathcal{FB}_{(1,1)}$ by performing derivatives w.r.t. relevant masses as same as other threshold functions mentioned above.




\bibliography{ref-lib}

\begin{thebibliography}{60}%
\makeatletter
\providecommand \@ifxundefined [1]{%
 \@ifx{#1\undefined}
}%
\providecommand \@ifnum [1]{%
 \ifnum #1\expandafter \@firstoftwo
 \else \expandafter \@secondoftwo
 \fi
}%
\providecommand \@ifx [1]{%
 \ifx #1\expandafter \@firstoftwo
 \else \expandafter \@secondoftwo
 \fi
}%
\providecommand \natexlab [1]{#1}%
\providecommand \enquote  [1]{``#1''}%
\providecommand \bibnamefont  [1]{#1}%
\providecommand \bibfnamefont [1]{#1}%
\providecommand \citenamefont [1]{#1}%
\providecommand \href@noop [0]{\@secondoftwo}%
\providecommand \href [0]{\begingroup \@sanitize@url \@href}%
\providecommand \@href[1]{\@@startlink{#1}\@@href}%
\providecommand \@@href[1]{\endgroup#1\@@endlink}%
\providecommand \@sanitize@url [0]{\catcode `\\12\catcode `\$12\catcode
  `\&12\catcode `\#12\catcode `\^12\catcode `\_12\catcode `\%12\relax}%
\providecommand \@@startlink[1]{}%
\providecommand \@@endlink[0]{}%
\providecommand \url  [0]{\begingroup\@sanitize@url \@url }%
\providecommand \@url [1]{\endgroup\@href {#1}{\urlprefix }}%
\providecommand \urlprefix  [0]{URL }%
\providecommand \Eprint [0]{\href }%
\providecommand \doibase [0]{http://dx.doi.org/}%
\providecommand \selectlanguage [0]{\@gobble}%
\providecommand \bibinfo  [0]{\@secondoftwo}%
\providecommand \bibfield  [0]{\@secondoftwo}%
\providecommand \translation [1]{[#1]}%
\providecommand \BibitemOpen [0]{}%
\providecommand \bibitemStop [0]{}%
\providecommand \bibitemNoStop [0]{.\EOS\space}%
\providecommand \EOS [0]{\spacefactor3000\relax}%
\providecommand \BibitemShut  [1]{\csname bibitem#1\endcsname}%
\let\auto@bib@innerbib\@empty
\bibitem [{\citenamefont {Adamczyk}\ \emph {et~al.}(2014)\citenamefont
  {Adamczyk} \emph {et~al.}}]{Adamczyk:2013dal}%
  \BibitemOpen
  \bibfield  {author} {\bibinfo {author} {\bibfnamefont {L.}~\bibnamefont
  {Adamczyk}} \emph {et~al.} (\bibinfo {collaboration} {STAR}),\ }\href
  {\doibase 10.1103/PhysRevLett.112.032302} {\bibfield  {journal} {\bibinfo
  {journal} {Phys. Rev. Lett.}\ }\textbf {\bibinfo {volume} {112}},\ \bibinfo
  {pages} {032302} (\bibinfo {year} {2014})},\ \Eprint
  {http://arxiv.org/abs/1309.5681} {arXiv:1309.5681 [nucl-ex]} \BibitemShut
  {NoStop}%
\bibitem [{\citenamefont {Luo}(2015)}]{Luo:2015ewa}%
  \BibitemOpen
  \bibfield  {author} {\bibinfo {author} {\bibfnamefont {X.}~\bibnamefont
  {Luo}} (\bibinfo {collaboration} {STAR}),\ }\bibfield  {booktitle} {\emph
  {\bibinfo {booktitle} {{Proceedings, 9th International Workshop on Critical
  Point and Onset of Deconfinement (CPOD 2014): Bielefeld, Germany, November
  17-21, 2014}}},\ }\href@noop {} {\bibfield  {journal} {\bibinfo  {journal}
  {PoS}\ }\textbf {\bibinfo {volume} {CPOD2014}},\ \bibinfo {pages} {019}
  (\bibinfo {year} {2015})},\ \Eprint {http://arxiv.org/abs/1503.02558}
  {arXiv:1503.02558 [nucl-ex]} \BibitemShut {NoStop}%
\bibitem [{\citenamefont {Luo}\ and\ \citenamefont {Xu}(2017)}]{Luo:2017faz}%
  \BibitemOpen
  \bibfield  {author} {\bibinfo {author} {\bibfnamefont {X.}~\bibnamefont
  {Luo}}\ and\ \bibinfo {author} {\bibfnamefont {N.}~\bibnamefont {Xu}},\
  }\href {\doibase 10.1007/s41365-017-0257-0} {\bibfield  {journal} {\bibinfo
  {journal} {Nucl. Sci. Tech.}\ }\textbf {\bibinfo {volume} {28}},\ \bibinfo
  {pages} {112} (\bibinfo {year} {2017})},\ \Eprint
  {http://arxiv.org/abs/1701.02105} {arXiv:1701.02105 [nucl-ex]} \BibitemShut
  {NoStop}%
\bibitem [{\citenamefont {Bellwied}\ \emph {et~al.}(2015)\citenamefont
  {Bellwied}, \citenamefont {Borsanyi}, \citenamefont {Fodor}, \citenamefont
  {Gnther}, \citenamefont {Katz}, \citenamefont {Ratti},\ and\ \citenamefont
  {Szabo}}]{Bellwied:2015rza}%
  \BibitemOpen
  \bibfield  {author} {\bibinfo {author} {\bibfnamefont {R.}~\bibnamefont
  {Bellwied}}, \bibinfo {author} {\bibfnamefont {S.}~\bibnamefont {Borsanyi}},
  \bibinfo {author} {\bibfnamefont {Z.}~\bibnamefont {Fodor}}, \bibinfo
  {author} {\bibfnamefont {J.}~\bibnamefont {Gnther}}, \bibinfo {author}
  {\bibfnamefont {S.~D.}\ \bibnamefont {Katz}}, \bibinfo {author}
  {\bibfnamefont {C.}~\bibnamefont {Ratti}}, \ and\ \bibinfo {author}
  {\bibfnamefont {K.~K.}\ \bibnamefont {Szabo}},\ }\href {\doibase
  10.1016/j.physletb.2015.11.011} {\bibfield  {journal} {\bibinfo  {journal}
  {Phys. Lett.}\ }\textbf {\bibinfo {volume} {B751}},\ \bibinfo {pages} {559}
  (\bibinfo {year} {2015})},\ \Eprint {http://arxiv.org/abs/1507.07510}
  {arXiv:1507.07510 [hep-lat]} \BibitemShut {NoStop}%
\bibitem [{\citenamefont {Bazavov}\ \emph {et~al.}(2019)\citenamefont {Bazavov}
  \emph {et~al.}}]{Bazavov:2018mes}%
  \BibitemOpen
  \bibfield  {author} {\bibinfo {author} {\bibfnamefont {A.}~\bibnamefont
  {Bazavov}} \emph {et~al.} (\bibinfo {collaboration} {HotQCD}),\ }\href
  {\doibase 10.1016/j.physletb.2019.05.013} {\bibfield  {journal} {\bibinfo
  {journal} {Phys. Lett.}\ }\textbf {\bibinfo {volume} {B795}},\ \bibinfo
  {pages} {15} (\bibinfo {year} {2019})},\ \Eprint
  {http://arxiv.org/abs/1812.08235} {arXiv:1812.08235 [hep-lat]} \BibitemShut
  {NoStop}%
\bibitem [{\citenamefont {Bazavov}\ \emph
  {et~al.}(2017{\natexlab{a}})\citenamefont {Bazavov} \emph
  {et~al.}}]{Bazavov:2017dus}%
  \BibitemOpen
  \bibfield  {author} {\bibinfo {author} {\bibfnamefont {A.}~\bibnamefont
  {Bazavov}} \emph {et~al.},\ }\href {\doibase 10.1103/PhysRevD.95.054504}
  {\bibfield  {journal} {\bibinfo  {journal} {Phys. Rev.}\ }\textbf {\bibinfo
  {volume} {D95}},\ \bibinfo {pages} {054504} (\bibinfo {year}
  {2017}{\natexlab{a}})},\ \Eprint {http://arxiv.org/abs/1701.04325}
  {arXiv:1701.04325 [hep-lat]} \BibitemShut {NoStop}%
\bibitem [{\citenamefont {Bazavov}\ \emph
  {et~al.}(2017{\natexlab{b}})\citenamefont {Bazavov} \emph
  {et~al.}}]{Bazavov:2017tot}%
  \BibitemOpen
  \bibfield  {author} {\bibinfo {author} {\bibfnamefont {A.}~\bibnamefont
  {Bazavov}} \emph {et~al.} (\bibinfo {collaboration} {HotQCD}),\ }\href
  {\doibase 10.1103/PhysRevD.96.074510} {\bibfield  {journal} {\bibinfo
  {journal} {Phys. Rev.}\ }\textbf {\bibinfo {volume} {D96}},\ \bibinfo {pages}
  {074510} (\bibinfo {year} {2017}{\natexlab{b}})},\ \Eprint
  {http://arxiv.org/abs/1708.04897} {arXiv:1708.04897 [hep-lat]} \BibitemShut
  {NoStop}%
\bibitem [{\citenamefont {Borsanyi}\ \emph {et~al.}(2018)\citenamefont
  {Borsanyi}, \citenamefont {Fodor}, \citenamefont {Guenther}, \citenamefont
  {Katz}, \citenamefont {Szabo}, \citenamefont {Pasztor}, \citenamefont
  {Portillo},\ and\ \citenamefont {Ratti}}]{Borsanyi:2018grb}%
  \BibitemOpen
  \bibfield  {author} {\bibinfo {author} {\bibfnamefont {S.}~\bibnamefont
  {Borsanyi}}, \bibinfo {author} {\bibfnamefont {Z.}~\bibnamefont {Fodor}},
  \bibinfo {author} {\bibfnamefont {J.~N.}\ \bibnamefont {Guenther}}, \bibinfo
  {author} {\bibfnamefont {S.~K.}\ \bibnamefont {Katz}}, \bibinfo {author}
  {\bibfnamefont {K.~K.}\ \bibnamefont {Szabo}}, \bibinfo {author}
  {\bibfnamefont {A.}~\bibnamefont {Pasztor}}, \bibinfo {author} {\bibfnamefont
  {I.}~\bibnamefont {Portillo}}, \ and\ \bibinfo {author} {\bibfnamefont
  {C.}~\bibnamefont {Ratti}},\ }\href {\doibase 10.1007/JHEP10(2018)205}
  {\bibfield  {journal} {\bibinfo  {journal} {JHEP}\ }\textbf {\bibinfo
  {volume} {10}},\ \bibinfo {pages} {205} (\bibinfo {year} {2018})},\ \Eprint
  {http://arxiv.org/abs/1805.04445} {arXiv:1805.04445 [hep-lat]} \BibitemShut
  {NoStop}%
\bibitem [{\citenamefont {Ding}\ \emph {et~al.}(2019)\citenamefont {Ding} \emph
  {et~al.}}]{Ding:2019prx}%
  \BibitemOpen
  \bibfield  {author} {\bibinfo {author} {\bibfnamefont {H.~T.}\ \bibnamefont
  {Ding}} \emph {et~al.},\ }\href@noop {} {\  (\bibinfo {year} {2019})},\
  \Eprint {http://arxiv.org/abs/1903.04801} {arXiv:1903.04801 [hep-lat]}
  \BibitemShut {NoStop}%
\bibitem [{\citenamefont {Fischer}(2019)}]{Fischer:2018sdj}%
  \BibitemOpen
  \bibfield  {author} {\bibinfo {author} {\bibfnamefont {C.~S.}\ \bibnamefont
  {Fischer}},\ }\href {\doibase 10.1016/j.ppnp.2019.01.002} {\bibfield
  {journal} {\bibinfo  {journal} {Prog. Part. Nucl. Phys.}\ }\textbf {\bibinfo
  {volume} {105}},\ \bibinfo {pages} {1} (\bibinfo {year} {2019})},\ \Eprint
  {http://arxiv.org/abs/1810.12938} {arXiv:1810.12938 [hep-ph]} \BibitemShut
  {NoStop}%
\bibitem [{\citenamefont {Pawlowski}(2007)}]{Pawlowski:2005xe}%
  \BibitemOpen
  \bibfield  {author} {\bibinfo {author} {\bibfnamefont {J.~M.}\ \bibnamefont
  {Pawlowski}},\ }\href {\doibase 10.1016/j.aop.2007.01.007} {\bibfield
  {journal} {\bibinfo  {journal} {Annals Phys.}\ }\textbf {\bibinfo {volume}
  {322}},\ \bibinfo {pages} {2831} (\bibinfo {year} {2007})},\ \Eprint
  {http://arxiv.org/abs/hep-th/0512261} {arXiv:hep-th/0512261 [hep-th]}
  \BibitemShut {NoStop}%
\bibitem [{\citenamefont {Schaefer}\ and\ \citenamefont
  {Wambach}(2005)}]{Schaefer:2004en}%
  \BibitemOpen
  \bibfield  {author} {\bibinfo {author} {\bibfnamefont {B.-J.}\ \bibnamefont
  {Schaefer}}\ and\ \bibinfo {author} {\bibfnamefont {J.}~\bibnamefont
  {Wambach}},\ }\href {\doibase 10.1016/j.nuclphysa.2005.04.012} {\bibfield
  {journal} {\bibinfo  {journal} {Nucl. Phys.}\ }\textbf {\bibinfo {volume}
  {A757}},\ \bibinfo {pages} {479} (\bibinfo {year} {2005})},\ \Eprint
  {http://arxiv.org/abs/nucl-th/0403039} {arXiv:nucl-th/0403039 [nucl-th]}
  \BibitemShut {NoStop}%
\bibitem [{\citenamefont {Herbst}\ \emph {et~al.}(2013)\citenamefont {Herbst},
  \citenamefont {Pawlowski},\ and\ \citenamefont {Schaefer}}]{Herbst:2013ail}%
  \BibitemOpen
  \bibfield  {author} {\bibinfo {author} {\bibfnamefont {T.~K.}\ \bibnamefont
  {Herbst}}, \bibinfo {author} {\bibfnamefont {J.~M.}\ \bibnamefont
  {Pawlowski}}, \ and\ \bibinfo {author} {\bibfnamefont {B.-J.}\ \bibnamefont
  {Schaefer}},\ }\href {\doibase 10.1103/PhysRevD.88.014007} {\bibfield
  {journal} {\bibinfo  {journal} {Phys. Rev.}\ }\textbf {\bibinfo {volume}
  {D88}},\ \bibinfo {pages} {014007} (\bibinfo {year} {2013})},\ \Eprint
  {http://arxiv.org/abs/1302.1426} {arXiv:1302.1426 [hep-ph]} \BibitemShut
  {NoStop}%
\bibitem [{\citenamefont {Braun}\ \emph {et~al.}(2011)\citenamefont {Braun},
  \citenamefont {Haas}, \citenamefont {Marhauser},\ and\ \citenamefont
  {Pawlowski}}]{Braun:2009gm}%
  \BibitemOpen
  \bibfield  {author} {\bibinfo {author} {\bibfnamefont {J.}~\bibnamefont
  {Braun}}, \bibinfo {author} {\bibfnamefont {L.~M.}\ \bibnamefont {Haas}},
  \bibinfo {author} {\bibfnamefont {F.}~\bibnamefont {Marhauser}}, \ and\
  \bibinfo {author} {\bibfnamefont {J.~M.}\ \bibnamefont {Pawlowski}},\ }\href
  {\doibase 10.1103/PhysRevLett.106.022002} {\bibfield  {journal} {\bibinfo
  {journal} {Phys. Rev. Lett.}\ }\textbf {\bibinfo {volume} {106}},\ \bibinfo
  {pages} {022002} (\bibinfo {year} {2011})},\ \Eprint
  {http://arxiv.org/abs/0908.0008} {arXiv:0908.0008 [hep-ph]} \BibitemShut
  {NoStop}%
\bibitem [{\citenamefont {Fischer}\ \emph {et~al.}(2014)\citenamefont
  {Fischer}, \citenamefont {Luecker},\ and\ \citenamefont
  {Welzbacher}}]{Fischer:2014ata}%
  \BibitemOpen
  \bibfield  {author} {\bibinfo {author} {\bibfnamefont {C.~S.}\ \bibnamefont
  {Fischer}}, \bibinfo {author} {\bibfnamefont {J.}~\bibnamefont {Luecker}}, \
  and\ \bibinfo {author} {\bibfnamefont {C.~A.}\ \bibnamefont {Welzbacher}},\
  }\href {\doibase 10.1103/PhysRevD.90.034022} {\bibfield  {journal} {\bibinfo
  {journal} {Phys. Rev.}\ }\textbf {\bibinfo {volume} {D90}},\ \bibinfo {pages}
  {034022} (\bibinfo {year} {2014})},\ \Eprint {http://arxiv.org/abs/1405.4762}
  {arXiv:1405.4762 [hep-ph]} \BibitemShut {NoStop}%
\bibitem [{\citenamefont {Gao}\ \emph {et~al.}(2016)\citenamefont {Gao},
  \citenamefont {Chen}, \citenamefont {Liu}, \citenamefont {Qin}, \citenamefont
  {Roberts},\ and\ \citenamefont {Schmidt}}]{Gao:2015kea}%
  \BibitemOpen
  \bibfield  {author} {\bibinfo {author} {\bibfnamefont {F.}~\bibnamefont
  {Gao}}, \bibinfo {author} {\bibfnamefont {J.}~\bibnamefont {Chen}}, \bibinfo
  {author} {\bibfnamefont {Y.-X.}\ \bibnamefont {Liu}}, \bibinfo {author}
  {\bibfnamefont {S.-X.}\ \bibnamefont {Qin}}, \bibinfo {author} {\bibfnamefont
  {C.~D.}\ \bibnamefont {Roberts}}, \ and\ \bibinfo {author} {\bibfnamefont
  {S.~M.}\ \bibnamefont {Schmidt}},\ }\href {\doibase
  10.1103/PhysRevD.93.094019} {\bibfield  {journal} {\bibinfo  {journal} {Phys.
  Rev.}\ }\textbf {\bibinfo {volume} {D93}},\ \bibinfo {pages} {094019}
  (\bibinfo {year} {2016})},\ \Eprint {http://arxiv.org/abs/1507.00875}
  {arXiv:1507.00875 [nucl-th]} \BibitemShut {NoStop}%
\bibitem [{\citenamefont {Isserstedt}\ \emph {et~al.}(2019)\citenamefont
  {Isserstedt}, \citenamefont {Buballa}, \citenamefont {Fischer},\ and\
  \citenamefont {Gunkel}}]{Isserstedt:2019pgx}%
  \BibitemOpen
  \bibfield  {author} {\bibinfo {author} {\bibfnamefont {P.}~\bibnamefont
  {Isserstedt}}, \bibinfo {author} {\bibfnamefont {M.}~\bibnamefont {Buballa}},
  \bibinfo {author} {\bibfnamefont {C.~S.}\ \bibnamefont {Fischer}}, \ and\
  \bibinfo {author} {\bibfnamefont {P.~J.}\ \bibnamefont {Gunkel}},\
  }\href@noop {} {\  (\bibinfo {year} {2019})},\ \Eprint
  {http://arxiv.org/abs/1906.11644} {arXiv:1906.11644 [hep-ph]} \BibitemShut
  {NoStop}%
\bibitem [{\citenamefont {Fu}\ \emph {et~al.}(2019)\citenamefont {Fu},
  \citenamefont {Pawlowski},\ and\ \citenamefont {Rennecke}}]{Fu:2019hdw}%
  \BibitemOpen
  \bibfield  {author} {\bibinfo {author} {\bibfnamefont {W.-j.}\ \bibnamefont
  {Fu}}, \bibinfo {author} {\bibfnamefont {J.~M.}\ \bibnamefont {Pawlowski}}, \
  and\ \bibinfo {author} {\bibfnamefont {F.}~\bibnamefont {Rennecke}},\
  }\href@noop {} {\  (\bibinfo {year} {2019})},\ \Eprint
  {http://arxiv.org/abs/1909.02991} {arXiv:1909.02991 [hep-ph]} \BibitemShut
  {NoStop}%
\bibitem [{\citenamefont {Berges}\ \emph {et~al.}(2002)\citenamefont {Berges},
  \citenamefont {Tetradis},\ and\ \citenamefont {Wetterich}}]{Berges:2000ew}%
  \BibitemOpen
  \bibfield  {author} {\bibinfo {author} {\bibfnamefont {J.}~\bibnamefont
  {Berges}}, \bibinfo {author} {\bibfnamefont {N.}~\bibnamefont {Tetradis}}, \
  and\ \bibinfo {author} {\bibfnamefont {C.}~\bibnamefont {Wetterich}},\ }\href
  {\doibase 10.1016/S0370-1573(01)00098-9} {\bibfield  {journal} {\bibinfo
  {journal} {Phys. Rept.}\ }\textbf {\bibinfo {volume} {363}},\ \bibinfo
  {pages} {223} (\bibinfo {year} {2002})},\ \Eprint
  {http://arxiv.org/abs/hep-ph/0005122} {arXiv:hep-ph/0005122 [hep-ph]}
  \BibitemShut {NoStop}%
\bibitem [{\citenamefont {Pawlowski}(2011)}]{Pawlowski:2010ht}%
  \BibitemOpen
  \bibfield  {author} {\bibinfo {author} {\bibfnamefont {J.~M.}\ \bibnamefont
  {Pawlowski}},\ }\bibfield  {booktitle} {\emph {\bibinfo {booktitle}
  {{Proceedings, 9th Conference on Quark Confinement and the Hadron Spectrum:
  Madrid, Spain, 30 Aug-3 Sep 2010}}},\ }\href {\doibase 10.1063/1.3574945}
  {\bibfield  {journal} {\bibinfo  {journal} {AIP Conf. Proc.}\ }\textbf
  {\bibinfo {volume} {1343}},\ \bibinfo {pages} {75} (\bibinfo {year}
  {2011})},\ \Eprint {http://arxiv.org/abs/1012.5075} {arXiv:1012.5075
  [hep-ph]} \BibitemShut {NoStop}%
\bibitem [{\citenamefont {Braun}(2012)}]{Braun:2011pp}%
  \BibitemOpen
  \bibfield  {author} {\bibinfo {author} {\bibfnamefont {J.}~\bibnamefont
  {Braun}},\ }\href {\doibase 10.1088/0954-3899/39/3/033001} {\bibfield
  {journal} {\bibinfo  {journal} {J. Phys.}\ }\textbf {\bibinfo {volume}
  {G39}},\ \bibinfo {pages} {033001} (\bibinfo {year} {2012})},\ \Eprint
  {http://arxiv.org/abs/1108.4449} {arXiv:1108.4449 [hep-ph]} \BibitemShut
  {NoStop}%
\bibitem [{\citenamefont {Mitter}\ \emph {et~al.}(2015)\citenamefont {Mitter},
  \citenamefont {Pawlowski},\ and\ \citenamefont
  {Strodthoff}}]{Mitter:2014wpa}%
  \BibitemOpen
  \bibfield  {author} {\bibinfo {author} {\bibfnamefont {M.}~\bibnamefont
  {Mitter}}, \bibinfo {author} {\bibfnamefont {J.~M.}\ \bibnamefont
  {Pawlowski}}, \ and\ \bibinfo {author} {\bibfnamefont {N.}~\bibnamefont
  {Strodthoff}},\ }\href {\doibase 10.1103/PhysRevD.91.054035} {\bibfield
  {journal} {\bibinfo  {journal} {Phys. Rev.}\ }\textbf {\bibinfo {volume}
  {D91}},\ \bibinfo {pages} {054035} (\bibinfo {year} {2015})},\ \Eprint
  {http://arxiv.org/abs/1411.7978} {arXiv:1411.7978 [hep-ph]} \BibitemShut
  {NoStop}%
\bibitem [{\citenamefont {Braun}\ \emph {et~al.}(2016)\citenamefont {Braun},
  \citenamefont {Fister}, \citenamefont {Pawlowski},\ and\ \citenamefont
  {Rennecke}}]{Braun:2014ata}%
  \BibitemOpen
  \bibfield  {author} {\bibinfo {author} {\bibfnamefont {J.}~\bibnamefont
  {Braun}}, \bibinfo {author} {\bibfnamefont {L.}~\bibnamefont {Fister}},
  \bibinfo {author} {\bibfnamefont {J.~M.}\ \bibnamefont {Pawlowski}}, \ and\
  \bibinfo {author} {\bibfnamefont {F.}~\bibnamefont {Rennecke}},\ }\href
  {\doibase 10.1103/PhysRevD.94.034016} {\bibfield  {journal} {\bibinfo
  {journal} {Phys. Rev.}\ }\textbf {\bibinfo {volume} {D94}},\ \bibinfo {pages}
  {034016} (\bibinfo {year} {2016})},\ \Eprint {http://arxiv.org/abs/1412.1045}
  {arXiv:1412.1045 [hep-ph]} \BibitemShut {NoStop}%
\bibitem [{\citenamefont {Rennecke}(2015)}]{Rennecke:2015eba}%
  \BibitemOpen
  \bibfield  {author} {\bibinfo {author} {\bibfnamefont {F.}~\bibnamefont
  {Rennecke}},\ }\href {\doibase 10.1103/PhysRevD.92.076012} {\bibfield
  {journal} {\bibinfo  {journal} {Phys. Rev.}\ }\textbf {\bibinfo {volume}
  {D92}},\ \bibinfo {pages} {076012} (\bibinfo {year} {2015})},\ \Eprint
  {http://arxiv.org/abs/1504.03585} {arXiv:1504.03585 [hep-ph]} \BibitemShut
  {NoStop}%
\bibitem [{\citenamefont {Cyrol}\ \emph {et~al.}(2016)\citenamefont {Cyrol},
  \citenamefont {Fister}, \citenamefont {Mitter}, \citenamefont {Pawlowski},\
  and\ \citenamefont {Strodthoff}}]{Cyrol:2016tym}%
  \BibitemOpen
  \bibfield  {author} {\bibinfo {author} {\bibfnamefont {A.~K.}\ \bibnamefont
  {Cyrol}}, \bibinfo {author} {\bibfnamefont {L.}~\bibnamefont {Fister}},
  \bibinfo {author} {\bibfnamefont {M.}~\bibnamefont {Mitter}}, \bibinfo
  {author} {\bibfnamefont {J.~M.}\ \bibnamefont {Pawlowski}}, \ and\ \bibinfo
  {author} {\bibfnamefont {N.}~\bibnamefont {Strodthoff}},\ }\href {\doibase
  10.1103/PhysRevD.94.054005} {\bibfield  {journal} {\bibinfo  {journal} {Phys.
  Rev.}\ }\textbf {\bibinfo {volume} {D94}},\ \bibinfo {pages} {054005}
  (\bibinfo {year} {2016})},\ \Eprint {http://arxiv.org/abs/1605.01856}
  {arXiv:1605.01856 [hep-ph]} \BibitemShut {NoStop}%
\bibitem [{\citenamefont {Cyrol}\ \emph
  {et~al.}(2018{\natexlab{a}})\citenamefont {Cyrol}, \citenamefont {Mitter},
  \citenamefont {Pawlowski},\ and\ \citenamefont {Strodthoff}}]{Cyrol:2017ewj}%
  \BibitemOpen
  \bibfield  {author} {\bibinfo {author} {\bibfnamefont {A.~K.}\ \bibnamefont
  {Cyrol}}, \bibinfo {author} {\bibfnamefont {M.}~\bibnamefont {Mitter}},
  \bibinfo {author} {\bibfnamefont {J.~M.}\ \bibnamefont {Pawlowski}}, \ and\
  \bibinfo {author} {\bibfnamefont {N.}~\bibnamefont {Strodthoff}},\ }\href
  {\doibase 10.1103/PhysRevD.97.054006} {\bibfield  {journal} {\bibinfo
  {journal} {Phys. Rev.}\ }\textbf {\bibinfo {volume} {D97}},\ \bibinfo {pages}
  {054006} (\bibinfo {year} {2018}{\natexlab{a}})},\ \Eprint
  {http://arxiv.org/abs/1706.06326} {arXiv:1706.06326 [hep-ph]} \BibitemShut
  {NoStop}%
\bibitem [{\citenamefont {Cyrol}\ \emph
  {et~al.}(2018{\natexlab{b}})\citenamefont {Cyrol}, \citenamefont {Mitter},
  \citenamefont {Pawlowski},\ and\ \citenamefont {Strodthoff}}]{Cyrol:2017qkl}%
  \BibitemOpen
  \bibfield  {author} {\bibinfo {author} {\bibfnamefont {A.~K.}\ \bibnamefont
  {Cyrol}}, \bibinfo {author} {\bibfnamefont {M.}~\bibnamefont {Mitter}},
  \bibinfo {author} {\bibfnamefont {J.~M.}\ \bibnamefont {Pawlowski}}, \ and\
  \bibinfo {author} {\bibfnamefont {N.}~\bibnamefont {Strodthoff}},\ }\href
  {\doibase 10.1103/PhysRevD.97.054015} {\bibfield  {journal} {\bibinfo
  {journal} {Phys. Rev.}\ }\textbf {\bibinfo {volume} {D97}},\ \bibinfo {pages}
  {054015} (\bibinfo {year} {2018}{\natexlab{b}})},\ \Eprint
  {http://arxiv.org/abs/1708.03482} {arXiv:1708.03482 [hep-ph]} \BibitemShut
  {NoStop}%
\bibitem [{\citenamefont {Schaefer}\ and\ \citenamefont
  {Wambach}(2008)}]{Schaefer:2006sr}%
  \BibitemOpen
  \bibfield  {author} {\bibinfo {author} {\bibfnamefont {B.-J.}\ \bibnamefont
  {Schaefer}}\ and\ \bibinfo {author} {\bibfnamefont {J.}~\bibnamefont
  {Wambach}},\ }\bibfield  {booktitle} {\emph {\bibinfo {booktitle} {{Helmholtz
  International Summer School on Dense Matter in Heavy Ion Collisions and
  Astrophysics Dubna, Russia, August 21-September 1, 2006}}},\ }\href {\doibase
  10.1134/S1063779608070083} {\bibfield  {journal} {\bibinfo  {journal} {Phys.
  Part. Nucl.}\ }\textbf {\bibinfo {volume} {39}},\ \bibinfo {pages} {1025}
  (\bibinfo {year} {2008})},\ \Eprint {http://arxiv.org/abs/hep-ph/0611191}
  {arXiv:hep-ph/0611191 [hep-ph]} \BibitemShut {NoStop}%
\bibitem [{\citenamefont {Skokov}\ \emph {et~al.}(2010)\citenamefont {Skokov},
  \citenamefont {Stokic}, \citenamefont {Friman},\ and\ \citenamefont
  {Redlich}}]{Skokov:2010wb}%
  \BibitemOpen
  \bibfield  {author} {\bibinfo {author} {\bibfnamefont {V.}~\bibnamefont
  {Skokov}}, \bibinfo {author} {\bibfnamefont {B.}~\bibnamefont {Stokic}},
  \bibinfo {author} {\bibfnamefont {B.}~\bibnamefont {Friman}}, \ and\ \bibinfo
  {author} {\bibfnamefont {K.}~\bibnamefont {Redlich}},\ }\href {\doibase
  10.1103/PhysRevC.82.015206} {\bibfield  {journal} {\bibinfo  {journal} {Phys.
  Rev.}\ }\textbf {\bibinfo {volume} {C82}},\ \bibinfo {pages} {015206}
  (\bibinfo {year} {2010})},\ \Eprint {http://arxiv.org/abs/1004.2665}
  {arXiv:1004.2665 [hep-ph]} \BibitemShut {NoStop}%
\bibitem [{\citenamefont {Herbst}\ \emph {et~al.}(2011)\citenamefont {Herbst},
  \citenamefont {Pawlowski},\ and\ \citenamefont {Schaefer}}]{Herbst:2010rf}%
  \BibitemOpen
  \bibfield  {author} {\bibinfo {author} {\bibfnamefont {T.~K.}\ \bibnamefont
  {Herbst}}, \bibinfo {author} {\bibfnamefont {J.~M.}\ \bibnamefont
  {Pawlowski}}, \ and\ \bibinfo {author} {\bibfnamefont {B.-J.}\ \bibnamefont
  {Schaefer}},\ }\href {\doibase 10.1016/j.physletb.2010.12.003} {\bibfield
  {journal} {\bibinfo  {journal} {Phys. Lett.}\ }\textbf {\bibinfo {volume}
  {B696}},\ \bibinfo {pages} {58} (\bibinfo {year} {2011})},\ \Eprint
  {http://arxiv.org/abs/1008.0081} {arXiv:1008.0081 [hep-ph]} \BibitemShut
  {NoStop}%
\bibitem [{\citenamefont {Schaefer}\ and\ \citenamefont
  {Wagner}(2012)}]{Schaefer:2011ex}%
  \BibitemOpen
  \bibfield  {author} {\bibinfo {author} {\bibfnamefont {B.~J.}\ \bibnamefont
  {Schaefer}}\ and\ \bibinfo {author} {\bibfnamefont {M.}~\bibnamefont
  {Wagner}},\ }\href {\doibase 10.1103/PhysRevD.85.034027} {\bibfield
  {journal} {\bibinfo  {journal} {Phys. Rev.}\ }\textbf {\bibinfo {volume}
  {D85}},\ \bibinfo {pages} {034027} (\bibinfo {year} {2012})},\ \Eprint
  {http://arxiv.org/abs/1111.6871} {arXiv:1111.6871 [hep-ph]} \BibitemShut
  {NoStop}%
\bibitem [{\citenamefont {Mitter}\ and\ \citenamefont
  {Schaefer}(2014)}]{Mitter:2013fxa}%
  \BibitemOpen
  \bibfield  {author} {\bibinfo {author} {\bibfnamefont {M.}~\bibnamefont
  {Mitter}}\ and\ \bibinfo {author} {\bibfnamefont {B.-J.}\ \bibnamefont
  {Schaefer}},\ }\href {\doibase 10.1103/PhysRevD.89.054027} {\bibfield
  {journal} {\bibinfo  {journal} {Phys. Rev.}\ }\textbf {\bibinfo {volume}
  {D89}},\ \bibinfo {pages} {054027} (\bibinfo {year} {2014})},\ \Eprint
  {http://arxiv.org/abs/1308.3176} {arXiv:1308.3176 [hep-ph]} \BibitemShut
  {NoStop}%
\bibitem [{\citenamefont {Herbst}\ \emph {et~al.}(2014)\citenamefont {Herbst},
  \citenamefont {Mitter}, \citenamefont {Pawlowski}, \citenamefont {Schaefer},\
  and\ \citenamefont {Stiele}}]{Herbst:2013ufa}%
  \BibitemOpen
  \bibfield  {author} {\bibinfo {author} {\bibfnamefont {T.~K.}\ \bibnamefont
  {Herbst}}, \bibinfo {author} {\bibfnamefont {M.}~\bibnamefont {Mitter}},
  \bibinfo {author} {\bibfnamefont {J.~M.}\ \bibnamefont {Pawlowski}}, \bibinfo
  {author} {\bibfnamefont {B.-J.}\ \bibnamefont {Schaefer}}, \ and\ \bibinfo
  {author} {\bibfnamefont {R.}~\bibnamefont {Stiele}},\ }\href {\doibase
  10.1016/j.physletb.2014.02.045} {\bibfield  {journal} {\bibinfo  {journal}
  {Phys. Lett.}\ }\textbf {\bibinfo {volume} {B731}},\ \bibinfo {pages} {248}
  (\bibinfo {year} {2014})},\ \Eprint {http://arxiv.org/abs/1308.3621}
  {arXiv:1308.3621 [hep-ph]} \BibitemShut {NoStop}%
\bibitem [{\citenamefont {Tripolt}\ \emph {et~al.}(2014)\citenamefont
  {Tripolt}, \citenamefont {Strodthoff}, \citenamefont {von Smekal},\ and\
  \citenamefont {Wambach}}]{Tripolt:2013jra}%
  \BibitemOpen
  \bibfield  {author} {\bibinfo {author} {\bibfnamefont {R.-A.}\ \bibnamefont
  {Tripolt}}, \bibinfo {author} {\bibfnamefont {N.}~\bibnamefont {Strodthoff}},
  \bibinfo {author} {\bibfnamefont {L.}~\bibnamefont {von Smekal}}, \ and\
  \bibinfo {author} {\bibfnamefont {J.}~\bibnamefont {Wambach}},\ }\href
  {\doibase 10.1103/PhysRevD.89.034010} {\bibfield  {journal} {\bibinfo
  {journal} {Phys. Rev.}\ }\textbf {\bibinfo {volume} {D89}},\ \bibinfo {pages}
  {034010} (\bibinfo {year} {2014})},\ \Eprint {http://arxiv.org/abs/1311.0630}
  {arXiv:1311.0630 [hep-ph]} \BibitemShut {NoStop}%
\bibitem [{\citenamefont {Fu}\ and\ \citenamefont
  {Pawlowski}(2015)}]{Fu:2015naa}%
  \BibitemOpen
  \bibfield  {author} {\bibinfo {author} {\bibfnamefont {W.-j.}\ \bibnamefont
  {Fu}}\ and\ \bibinfo {author} {\bibfnamefont {J.~M.}\ \bibnamefont
  {Pawlowski}},\ }\href {\doibase 10.1103/PhysRevD.92.116006} {\bibfield
  {journal} {\bibinfo  {journal} {Phys. Rev.}\ }\textbf {\bibinfo {volume}
  {D92}},\ \bibinfo {pages} {116006} (\bibinfo {year} {2015})},\ \Eprint
  {http://arxiv.org/abs/1508.06504} {arXiv:1508.06504 [hep-ph]} \BibitemShut
  {NoStop}%
\bibitem [{\citenamefont {Fu}\ and\ \citenamefont
  {Pawlowski}(2016)}]{Fu:2015amv}%
  \BibitemOpen
  \bibfield  {author} {\bibinfo {author} {\bibfnamefont {W.-j.}\ \bibnamefont
  {Fu}}\ and\ \bibinfo {author} {\bibfnamefont {J.~M.}\ \bibnamefont
  {Pawlowski}},\ }\href {\doibase 10.1103/PhysRevD.93.091501} {\bibfield
  {journal} {\bibinfo  {journal} {Phys. Rev.}\ }\textbf {\bibinfo {volume}
  {D93}},\ \bibinfo {pages} {091501} (\bibinfo {year} {2016})},\ \Eprint
  {http://arxiv.org/abs/1512.08461} {arXiv:1512.08461 [hep-ph]} \BibitemShut
  {NoStop}%
\bibitem [{\citenamefont {Fu}\ \emph {et~al.}(2016)\citenamefont {Fu},
  \citenamefont {Pawlowski}, \citenamefont {Rennecke},\ and\ \citenamefont
  {Schaefer}}]{Fu:2016tey}%
  \BibitemOpen
  \bibfield  {author} {\bibinfo {author} {\bibfnamefont {W.-j.}\ \bibnamefont
  {Fu}}, \bibinfo {author} {\bibfnamefont {J.~M.}\ \bibnamefont {Pawlowski}},
  \bibinfo {author} {\bibfnamefont {F.}~\bibnamefont {Rennecke}}, \ and\
  \bibinfo {author} {\bibfnamefont {B.-J.}\ \bibnamefont {Schaefer}},\ }\href
  {\doibase 10.1103/PhysRevD.94.116020} {\bibfield  {journal} {\bibinfo
  {journal} {Phys. Rev.}\ }\textbf {\bibinfo {volume} {D94}},\ \bibinfo {pages}
  {116020} (\bibinfo {year} {2016})},\ \Eprint
  {http://arxiv.org/abs/1608.04302} {arXiv:1608.04302 [hep-ph]} \BibitemShut
  {NoStop}%
\bibitem [{\citenamefont {Rennecke}\ and\ \citenamefont
  {Schaefer}(2017)}]{Rennecke:2016tkm}%
  \BibitemOpen
  \bibfield  {author} {\bibinfo {author} {\bibfnamefont {F.}~\bibnamefont
  {Rennecke}}\ and\ \bibinfo {author} {\bibfnamefont {B.-J.}\ \bibnamefont
  {Schaefer}},\ }\href {\doibase 10.1103/PhysRevD.96.016009} {\bibfield
  {journal} {\bibinfo  {journal} {Phys. Rev.}\ }\textbf {\bibinfo {volume}
  {D96}},\ \bibinfo {pages} {016009} (\bibinfo {year} {2017})},\ \Eprint
  {http://arxiv.org/abs/1610.08748} {arXiv:1610.08748 [hep-ph]} \BibitemShut
  {NoStop}%
\bibitem [{\citenamefont {Jung}\ \emph {et~al.}(2017)\citenamefont {Jung},
  \citenamefont {Rennecke}, \citenamefont {Tripolt}, \citenamefont {von
  Smekal},\ and\ \citenamefont {Wambach}}]{Jung:2016yxl}%
  \BibitemOpen
  \bibfield  {author} {\bibinfo {author} {\bibfnamefont {C.}~\bibnamefont
  {Jung}}, \bibinfo {author} {\bibfnamefont {F.}~\bibnamefont {Rennecke}},
  \bibinfo {author} {\bibfnamefont {R.-A.}\ \bibnamefont {Tripolt}}, \bibinfo
  {author} {\bibfnamefont {L.}~\bibnamefont {von Smekal}}, \ and\ \bibinfo
  {author} {\bibfnamefont {J.}~\bibnamefont {Wambach}},\ }\href {\doibase
  10.1103/PhysRevD.95.036020} {\bibfield  {journal} {\bibinfo  {journal} {Phys.
  Rev.}\ }\textbf {\bibinfo {volume} {D95}},\ \bibinfo {pages} {036020}
  (\bibinfo {year} {2017})},\ \Eprint {http://arxiv.org/abs/1610.08754}
  {arXiv:1610.08754 [hep-ph]} \BibitemShut {NoStop}%
\bibitem [{\citenamefont {Braun}\ \emph {et~al.}(2017)\citenamefont {Braun},
  \citenamefont {Leonhardt},\ and\ \citenamefont {Pospiech}}]{Braun:2017srn}%
  \BibitemOpen
  \bibfield  {author} {\bibinfo {author} {\bibfnamefont {J.}~\bibnamefont
  {Braun}}, \bibinfo {author} {\bibfnamefont {M.}~\bibnamefont {Leonhardt}}, \
  and\ \bibinfo {author} {\bibfnamefont {M.}~\bibnamefont {Pospiech}},\ }\href
  {\doibase 10.1103/PhysRevD.96.076003} {\bibfield  {journal} {\bibinfo
  {journal} {Phys. Rev.}\ }\textbf {\bibinfo {volume} {D96}},\ \bibinfo {pages}
  {076003} (\bibinfo {year} {2017})},\ \Eprint
  {http://arxiv.org/abs/1705.00074} {arXiv:1705.00074 [hep-ph]} \BibitemShut
  {NoStop}%
\bibitem [{\citenamefont {Fu}\ \emph {et~al.}(2018{\natexlab{a}})\citenamefont
  {Fu}, \citenamefont {Pawlowski},\ and\ \citenamefont
  {Rennecke}}]{Fu:2018qsk}%
  \BibitemOpen
  \bibfield  {author} {\bibinfo {author} {\bibfnamefont {W.-j.}\ \bibnamefont
  {Fu}}, \bibinfo {author} {\bibfnamefont {J.~M.}\ \bibnamefont {Pawlowski}}, \
  and\ \bibinfo {author} {\bibfnamefont {F.}~\bibnamefont {Rennecke}},\
  }\href@noop {} {\  (\bibinfo {year} {2018}{\natexlab{a}})},\ \Eprint
  {http://arxiv.org/abs/1808.00410} {arXiv:1808.00410 [hep-ph]} \BibitemShut
  {NoStop}%
\bibitem [{\citenamefont {Fu}\ \emph {et~al.}(2018{\natexlab{b}})\citenamefont
  {Fu}, \citenamefont {Pawlowski},\ and\ \citenamefont
  {Rennecke}}]{Fu:2018swz}%
  \BibitemOpen
  \bibfield  {author} {\bibinfo {author} {\bibfnamefont {W.-j.}\ \bibnamefont
  {Fu}}, \bibinfo {author} {\bibfnamefont {J.~M.}\ \bibnamefont {Pawlowski}}, \
  and\ \bibinfo {author} {\bibfnamefont {F.}~\bibnamefont {Rennecke}},\
  }\href@noop {} {\  (\bibinfo {year} {2018}{\natexlab{b}})},\ \Eprint
  {http://arxiv.org/abs/1809.01594} {arXiv:1809.01594 [hep-ph]} \BibitemShut
  {NoStop}%
\bibitem [{\citenamefont {Sun}\ \emph {et~al.}(2018)\citenamefont {Sun},
  \citenamefont {Wen},\ and\ \citenamefont {Fu}}]{Sun:2018ozp}%
  \BibitemOpen
  \bibfield  {author} {\bibinfo {author} {\bibfnamefont {K.-x.}\ \bibnamefont
  {Sun}}, \bibinfo {author} {\bibfnamefont {R.}~\bibnamefont {Wen}}, \ and\
  \bibinfo {author} {\bibfnamefont {W.-j.}\ \bibnamefont {Fu}},\ }\href
  {\doibase 10.1103/PhysRevD.98.074028} {\bibfield  {journal} {\bibinfo
  {journal} {Phys. Rev.}\ }\textbf {\bibinfo {volume} {D98}},\ \bibinfo {pages}
  {074028} (\bibinfo {year} {2018})},\ \Eprint
  {http://arxiv.org/abs/1805.12025} {arXiv:1805.12025 [hep-ph]} \BibitemShut
  {NoStop}%
\bibitem [{\citenamefont {Wen}\ \emph {et~al.}(2019)\citenamefont {Wen},
  \citenamefont {Huang},\ and\ \citenamefont {Fu}}]{Wen:2018nkn}%
  \BibitemOpen
  \bibfield  {author} {\bibinfo {author} {\bibfnamefont {R.}~\bibnamefont
  {Wen}}, \bibinfo {author} {\bibfnamefont {C.}~\bibnamefont {Huang}}, \ and\
  \bibinfo {author} {\bibfnamefont {W.-J.}\ \bibnamefont {Fu}},\ }\href
  {\doibase 10.1103/PhysRevD.99.094019} {\bibfield  {journal} {\bibinfo
  {journal} {Phys. Rev.}\ }\textbf {\bibinfo {volume} {D99}},\ \bibinfo {pages}
  {094019} (\bibinfo {year} {2019})},\ \Eprint
  {http://arxiv.org/abs/1809.04233} {arXiv:1809.04233 [hep-ph]} \BibitemShut
  {NoStop}%
\bibitem [{\citenamefont {Li}\ \emph {et~al.}(2019)\citenamefont {Li},
  \citenamefont {Fu},\ and\ \citenamefont {Liu}}]{Li:2019nzj}%
  \BibitemOpen
  \bibfield  {author} {\bibinfo {author} {\bibfnamefont {X.}~\bibnamefont
  {Li}}, \bibinfo {author} {\bibfnamefont {W.-J.}\ \bibnamefont {Fu}}, \ and\
  \bibinfo {author} {\bibfnamefont {Y.-X.}\ \bibnamefont {Liu}},\ }\href
  {\doibase 10.1103/PhysRevD.99.074029} {\bibfield  {journal} {\bibinfo
  {journal} {Phys. Rev.}\ }\textbf {\bibinfo {volume} {D99}},\ \bibinfo {pages}
  {074029} (\bibinfo {year} {2019})},\ \Eprint
  {http://arxiv.org/abs/1902.03866} {arXiv:1902.03866 [hep-ph]} \BibitemShut
  {NoStop}%
\bibitem [{\citenamefont {Wen}\ and\ \citenamefont {Fu}(2019)}]{Wen:2019ruz}%
  \BibitemOpen
  \bibfield  {author} {\bibinfo {author} {\bibfnamefont {R.}~\bibnamefont
  {Wen}}\ and\ \bibinfo {author} {\bibfnamefont {W.-j.}\ \bibnamefont {Fu}},\
  }\href@noop {} {\  (\bibinfo {year} {2019})},\ \Eprint
  {http://arxiv.org/abs/1909.12564} {arXiv:1909.12564 [hep-ph]} \BibitemShut
  {NoStop}%
\bibitem [{\citenamefont {Buballa}(2005)}]{Buballa:2003qv}%
  \BibitemOpen
  \bibfield  {author} {\bibinfo {author} {\bibfnamefont {M.}~\bibnamefont
  {Buballa}},\ }\href {\doibase 10.1016/j.physrep.2004.11.004} {\bibfield
  {journal} {\bibinfo  {journal} {Phys. Rept.}\ }\textbf {\bibinfo {volume}
  {407}},\ \bibinfo {pages} {205} (\bibinfo {year} {2005})},\ \Eprint
  {http://arxiv.org/abs/hep-ph/0402234} {arXiv:hep-ph/0402234 [hep-ph]}
  \BibitemShut {NoStop}%
\bibitem [{\citenamefont {Fukushima}(2004)}]{Fukushima:2003fw}%
  \BibitemOpen
  \bibfield  {author} {\bibinfo {author} {\bibfnamefont {K.}~\bibnamefont
  {Fukushima}},\ }\href {\doibase 10.1016/j.physletb.2004.04.027} {\bibfield
  {journal} {\bibinfo  {journal} {Phys. Lett.}\ }\textbf {\bibinfo {volume}
  {B591}},\ \bibinfo {pages} {277} (\bibinfo {year} {2004})},\ \Eprint
  {http://arxiv.org/abs/hep-ph/0310121} {arXiv:hep-ph/0310121 [hep-ph]}
  \BibitemShut {NoStop}%
\bibitem [{\citenamefont {Ratti}\ \emph {et~al.}(2006)\citenamefont {Ratti},
  \citenamefont {Thaler},\ and\ \citenamefont {Weise}}]{Ratti:2005jh}%
  \BibitemOpen
  \bibfield  {author} {\bibinfo {author} {\bibfnamefont {C.}~\bibnamefont
  {Ratti}}, \bibinfo {author} {\bibfnamefont {M.~A.}\ \bibnamefont {Thaler}}, \
  and\ \bibinfo {author} {\bibfnamefont {W.}~\bibnamefont {Weise}},\ }\href
  {\doibase 10.1103/PhysRevD.73.014019} {\bibfield  {journal} {\bibinfo
  {journal} {Phys. Rev.}\ }\textbf {\bibinfo {volume} {D73}},\ \bibinfo {pages}
  {014019} (\bibinfo {year} {2006})},\ \Eprint
  {http://arxiv.org/abs/hep-ph/0506234} {arXiv:hep-ph/0506234 [hep-ph]}
  \BibitemShut {NoStop}%
\bibitem [{\citenamefont {Fu}\ \emph {et~al.}(2008)\citenamefont {Fu},
  \citenamefont {Zhang},\ and\ \citenamefont {Liu}}]{Fu:2007xc}%
  \BibitemOpen
  \bibfield  {author} {\bibinfo {author} {\bibfnamefont {W.-j.}\ \bibnamefont
  {Fu}}, \bibinfo {author} {\bibfnamefont {Z.}~\bibnamefont {Zhang}}, \ and\
  \bibinfo {author} {\bibfnamefont {Y.-x.}\ \bibnamefont {Liu}},\ }\href
  {\doibase 10.1103/PhysRevD.77.014006} {\bibfield  {journal} {\bibinfo
  {journal} {Phys. Rev.}\ }\textbf {\bibinfo {volume} {D77}},\ \bibinfo {pages}
  {014006} (\bibinfo {year} {2008})},\ \Eprint {http://arxiv.org/abs/0711.0154}
  {arXiv:0711.0154 [hep-ph]} \BibitemShut {NoStop}%
\bibitem [{\citenamefont {Schaefer}\ \emph {et~al.}(2007)\citenamefont
  {Schaefer}, \citenamefont {Pawlowski},\ and\ \citenamefont
  {Wambach}}]{Schaefer:2007pw}%
  \BibitemOpen
  \bibfield  {author} {\bibinfo {author} {\bibfnamefont {B.-J.}\ \bibnamefont
  {Schaefer}}, \bibinfo {author} {\bibfnamefont {J.~M.}\ \bibnamefont
  {Pawlowski}}, \ and\ \bibinfo {author} {\bibfnamefont {J.}~\bibnamefont
  {Wambach}},\ }\href {\doibase 10.1103/PhysRevD.76.074023} {\bibfield
  {journal} {\bibinfo  {journal} {Phys. Rev.}\ }\textbf {\bibinfo {volume}
  {D76}},\ \bibinfo {pages} {074023} (\bibinfo {year} {2007})},\ \Eprint
  {http://arxiv.org/abs/0704.3234} {arXiv:0704.3234 [hep-ph]} \BibitemShut
  {NoStop}%
\bibitem [{\citenamefont {Fu}\ \emph {et~al.}(2010)\citenamefont {Fu},
  \citenamefont {Liu},\ and\ \citenamefont {Wu}}]{Fu:2009wy}%
  \BibitemOpen
  \bibfield  {author} {\bibinfo {author} {\bibfnamefont {W.-j.}\ \bibnamefont
  {Fu}}, \bibinfo {author} {\bibfnamefont {Y.-x.}\ \bibnamefont {Liu}}, \ and\
  \bibinfo {author} {\bibfnamefont {Y.-L.}\ \bibnamefont {Wu}},\ }\href
  {\doibase 10.1103/PhysRevD.81.014028} {\bibfield  {journal} {\bibinfo
  {journal} {Phys. Rev.}\ }\textbf {\bibinfo {volume} {D81}},\ \bibinfo {pages}
  {014028} (\bibinfo {year} {2010})},\ \Eprint {http://arxiv.org/abs/0910.5783}
  {arXiv:0910.5783 [hep-ph]} \BibitemShut {NoStop}%
\bibitem [{\citenamefont {Fu}\ and\ \citenamefont {Wu}(2010)}]{Fu:2010ay}%
  \BibitemOpen
  \bibfield  {author} {\bibinfo {author} {\bibfnamefont {W.-j.}\ \bibnamefont
  {Fu}}\ and\ \bibinfo {author} {\bibfnamefont {Y.-l.}\ \bibnamefont {Wu}},\
  }\href {\doibase 10.1103/PhysRevD.82.074013} {\bibfield  {journal} {\bibinfo
  {journal} {Phys. Rev.}\ }\textbf {\bibinfo {volume} {D82}},\ \bibinfo {pages}
  {074013} (\bibinfo {year} {2010})},\ \Eprint {http://arxiv.org/abs/1008.3684}
  {arXiv:1008.3684 [hep-ph]} \BibitemShut {NoStop}%
\bibitem [{\citenamefont {Pawlowski}\ and\ \citenamefont
  {Rennecke}(2014)}]{Pawlowski:2014zaa}%
  \BibitemOpen
  \bibfield  {author} {\bibinfo {author} {\bibfnamefont {J.~M.}\ \bibnamefont
  {Pawlowski}}\ and\ \bibinfo {author} {\bibfnamefont {F.}~\bibnamefont
  {Rennecke}},\ }\href {\doibase 10.1103/PhysRevD.90.076002} {\bibfield
  {journal} {\bibinfo  {journal} {Phys. Rev.}\ }\textbf {\bibinfo {volume}
  {D90}},\ \bibinfo {pages} {076002} (\bibinfo {year} {2014})},\ \Eprint
  {http://arxiv.org/abs/1403.1179} {arXiv:1403.1179 [hep-ph]} \BibitemShut
  {NoStop}%
\bibitem [{\citenamefont {Helmboldt}\ \emph {et~al.}(2015)\citenamefont
  {Helmboldt}, \citenamefont {Pawlowski},\ and\ \citenamefont
  {Strodthoff}}]{Helmboldt:2014iya}%
  \BibitemOpen
  \bibfield  {author} {\bibinfo {author} {\bibfnamefont {A.~J.}\ \bibnamefont
  {Helmboldt}}, \bibinfo {author} {\bibfnamefont {J.~M.}\ \bibnamefont
  {Pawlowski}}, \ and\ \bibinfo {author} {\bibfnamefont {N.}~\bibnamefont
  {Strodthoff}},\ }\href {\doibase 10.1103/PhysRevD.91.054010} {\bibfield
  {journal} {\bibinfo  {journal} {Phys. Rev.}\ }\textbf {\bibinfo {volume}
  {D91}},\ \bibinfo {pages} {054010} (\bibinfo {year} {2015})},\ \Eprint
  {http://arxiv.org/abs/1409.8414} {arXiv:1409.8414 [hep-ph]} \BibitemShut
  {NoStop}%
\bibitem [{\citenamefont {Wetterich}(1993)}]{Wetterich:1992yh}%
  \BibitemOpen
  \bibfield  {author} {\bibinfo {author} {\bibfnamefont {C.}~\bibnamefont
  {Wetterich}},\ }\href {\doibase 10.1016/0370-2693(93)90726-X} {\bibfield
  {journal} {\bibinfo  {journal} {Phys. Lett.}\ }\textbf {\bibinfo {volume}
  {B301}},\ \bibinfo {pages} {90} (\bibinfo {year} {1993})}\BibitemShut
  {NoStop}%
\bibitem [{\citenamefont {Lo}\ \emph {et~al.}(2013)\citenamefont {Lo},
  \citenamefont {Friman}, \citenamefont {Kaczmarek}, \citenamefont {Redlich},\
  and\ \citenamefont {Sasaki}}]{Lo:2013hla}%
  \BibitemOpen
  \bibfield  {author} {\bibinfo {author} {\bibfnamefont {P.~M.}\ \bibnamefont
  {Lo}}, \bibinfo {author} {\bibfnamefont {B.}~\bibnamefont {Friman}}, \bibinfo
  {author} {\bibfnamefont {O.}~\bibnamefont {Kaczmarek}}, \bibinfo {author}
  {\bibfnamefont {K.}~\bibnamefont {Redlich}}, \ and\ \bibinfo {author}
  {\bibfnamefont {C.}~\bibnamefont {Sasaki}},\ }\href {\doibase
  10.1103/PhysRevD.88.074502} {\bibfield  {journal} {\bibinfo  {journal} {Phys.
  Rev.}\ }\textbf {\bibinfo {volume} {D88}},\ \bibinfo {pages} {074502}
  (\bibinfo {year} {2013})},\ \Eprint {http://arxiv.org/abs/1307.5958}
  {arXiv:1307.5958 [hep-lat]} \BibitemShut {NoStop}%
\bibitem [{\citenamefont {Haas}\ \emph {et~al.}(2013)\citenamefont {Haas},
  \citenamefont {Stiele}, \citenamefont {Braun}, \citenamefont {Pawlowski},\
  and\ \citenamefont {Schaffner-Bielich}}]{Haas:2013qwp}%
  \BibitemOpen
  \bibfield  {author} {\bibinfo {author} {\bibfnamefont {L.~M.}\ \bibnamefont
  {Haas}}, \bibinfo {author} {\bibfnamefont {R.}~\bibnamefont {Stiele}},
  \bibinfo {author} {\bibfnamefont {J.}~\bibnamefont {Braun}}, \bibinfo
  {author} {\bibfnamefont {J.~M.}\ \bibnamefont {Pawlowski}}, \ and\ \bibinfo
  {author} {\bibfnamefont {J.}~\bibnamefont {Schaffner-Bielich}},\ }\href
  {\doibase 10.1103/PhysRevD.87.076004} {\bibfield  {journal} {\bibinfo
  {journal} {Phys. Rev.}\ }\textbf {\bibinfo {volume} {D87}},\ \bibinfo {pages}
  {076004} (\bibinfo {year} {2013})},\ \Eprint {http://arxiv.org/abs/1302.1993}
  {arXiv:1302.1993 [hep-ph]} \BibitemShut {NoStop}%
\bibitem [{\citenamefont {Litim}(2000)}]{Litim:2000ci}%
  \BibitemOpen
  \bibfield  {author} {\bibinfo {author} {\bibfnamefont {D.~F.}\ \bibnamefont
  {Litim}},\ }\href {\doibase 10.1016/S0370-2693(00)00748-6} {\bibfield
  {journal} {\bibinfo  {journal} {Phys. Lett.}\ }\textbf {\bibinfo {volume}
  {B486}},\ \bibinfo {pages} {92} (\bibinfo {year} {2000})},\ \Eprint
  {http://arxiv.org/abs/hep-th/0005245} {arXiv:hep-th/0005245 [hep-th]}
  \BibitemShut {NoStop}%
\bibitem [{\citenamefont {Litim}(2001)}]{Litim:2001up}%
  \BibitemOpen
  \bibfield  {author} {\bibinfo {author} {\bibfnamefont {D.~F.}\ \bibnamefont
  {Litim}},\ }\href {\doibase 10.1103/PhysRevD.64.105007} {\bibfield  {journal}
  {\bibinfo  {journal} {Phys. Rev.}\ }\textbf {\bibinfo {volume} {D64}},\
  \bibinfo {pages} {105007} (\bibinfo {year} {2001})},\ \Eprint
  {http://arxiv.org/abs/hep-th/0103195} {arXiv:hep-th/0103195 [hep-th]}
  \BibitemShut {NoStop}%
\end{thebibliography}%

\end{document}